\documentclass[12pt]{article}
\makeatletter \@addtoreset{equation}{section} \makeatother
\renewcommand{\theequation}{\thesection.\arabic{equation}}
\newcommand{\ba}{\begin{array}}
\newcommand{\ea}{\end{array}}
\newcommand{\beq}{\begin{equation}}
\newcommand{\eeq}{\end{equation}}
\newcommand{\bea}{\begin{eqnarray}}
\newcommand{\eea}{\end{eqnarray}}

\def\bce{\begin{center}}
\def\ece{\end{center}}

\def\nonu{\nonumber}

\def\pa{\partial}

\def\be{\beta}

\def\th{\theta}

\def\la{\lambda}

\def\eps6{{\displaystyle \mathop{\epsilon}^{6}}{}}
\def\g6{{\displaystyle \mathop{g}^{6}}{}}

\def\nab6{{\displaystyle \mathop{\nabla}^{6}}{}}

\def\0{{\sst{(0)}}}
\def\1{{\sst{(1)}}}
\def\2{{\sst{(2)}}}
\def\3{{\sst{(3)}}}
\def\4{{\sst{(4)}}}
\def\5{{\sst{(5)}}}
\def\6{{\sst{(6)}}}
\def\7{{\sst{(7)}}}
\def\8{{\sst{(8)}}}

\def\ba{\begin{array}}
\def\ea{\end{array}}
\def\beq{\begin{equation}}
\def\eeq{\end{equation}}
\def\be{\begin{equation}}
\def\ee{\end{equation}}

\def\la{\lambda}
\def\eps{\epsilon}

\def\th{{\theta}}
\def\bth{{\overline{\theta}}}

\def\ba{\begin{array}}
\def\ea{\end{array}}
\def\beq{\begin{equation}}
\def\eeq{\end{equation}}
\def\be{\begin{equation}}
\def\ee{\end{equation}}

\def\la{\lambda}
\def\eps{\epsilon}

\def\th{{\theta}}
\def\bth{{\overline{\theta}}}

\def\eps6{{\displaystyle \mathop{\epsilon}^{6}}{}}

\def\nab6{{\displaystyle \mathop{\nabla}^{6}}{}}

\newcommand{\tz}{\frac{\theta_{12}}{z_{12}}}
\newcommand{\tzb}{\frac{\bar{\theta}_{12}}{z_{12}}}
\newcommand{\tzzbb}{\frac{\theta_{12} \bar{\theta}_{12}}{z_{12}^2}}
\newcommand{\tzbb}{\frac{\theta_{12} \bar{\theta}_{12}}{z_{12}}}

\newcommand{\bean}{\begin{eqnarray*}}
\newcommand{\eean}{\end{eqnarray*}}

\begin{document}
\thispagestyle{empty} \addtocounter{page}{-1}
   \begin{flushright}
\end{flushright}

\vspace*{1.3cm}
  
\centerline{ \Large \bf   
The Large $N$ 't Hooft Limit of   
Kazama-Suzuki Model}
\vspace*{1.5cm}
\centerline{{\bf Changhyun Ahn 
}} 
\vspace*{1.0cm} 
\centerline{\it 
Department of Physics, Kyungpook National University, Taegu
702-701, Korea} 
\vspace*{0.8cm} 
\centerline{\tt ahn@knu.ac.kr 
} 
\vskip2cm

\centerline{\bf Abstract}
\vspace*{0.5cm}

We consider ${\cal N}=2$ Kazama-Suzuki model on ${\bf
CP}^N=\frac{SU(N+1)}{SU(N) \times U(1)}$. It is known that
the ${\cal N}=2$ current algebra for the supersymmetric WZW model, at
level $k$, 
is a nonlinear algebra. The ${\cal N}=2$ ${\cal W}_3$ algebra
corresponding to $N=2$ was
recovered from the generalized GKO coset construction previously.  
For $N=4$, we construct one of the higher spin currents, in ${\cal
N}=2$ ${\cal W}_5$ algebra, with spins $(2, \frac{5}{2}, \frac{5}{2}, 3)$. 
The self-coupling constant 
in the operator product expansion of this current and itself 
depends on $N$ as well as $k$ explicitly.
We also observe a new higher spin primary current of spins $(3,
\frac{7}{2}, \frac{7}{2}, 4)$. 
From the behaviors of $N=2, 4$ cases, we expect the operator product
expansion of the lowest higher spin current and itself in
${\cal N}=2$ ${\cal W}_{N+1}$ algebra.
By taking the large $(N, k)$ limit on the various operator product 
expansions in components, we reproduce, at the linear order,
the corresponding operator product expansions in 
${\cal N}=2$ classical ${\cal W}_{\infty}^{\rm{cl}}[\lambda]$ algebra
which is the
asymptotic symmetry of the higher spin $AdS_3$ supergravity found recently.
 
\baselineskip=18pt
\newpage
\renewcommand{\theequation}
{\arabic{section}\mbox{.}\arabic{equation}}

\section{Introduction}

The duality between the $W_N$ minimal model conformal field theories and 
the higher spin theory of Vasiliev on the $AdS_3$ 
has been proposed by Gaberdiel and Gopakumar in \cite{GG}.
Very recently, in \cite{GG1},
this proposal has been clarified further and they  
claim that
the $W_N$ minimal model conformal field theory
is dual, in the 't Hooft $\frac{1}{N}$ expansion, 
to  the higher spin theory coupled to one complex scalar.
The duality can hold at finite $N$ because of the nontrivial
truncation of the quantum algebra of the higher spin theory.

In \cite{CHR}, the ${\cal N}=2$ supersymmetric extension of \cite{GG}, 
the higher spin $AdS_3$ supergravity,
has been studied where the dual conformal field theory is given by 
${\cal N}=2$ ${\bf CP}^N$ Kazama-Suzuki(KS) model in two dimensions. 
The supergravity partition function is computed and agrees with the
partition function from the superconformal field theory side.   
Moreover, this superconformal partition function in the KS
model in the 't Hooft limit is described, in detail, in \cite{CG}.  
Recently, 
in \cite{HP}, the asymptotic symmetry of the higher
spin $AdS_3$ supergravity is obtained, by following the work of
\cite{GH},
 and one of the nontrivial
checks for the duality \cite{CHR} is to identify the operator product expansions
between the lower higher spin currents in the KS model, 
in the 't Hooft limit, 
with the corresponding algebra in the classical ${\cal N}=2$  ${\cal
  W}_{\infty}^{\rm{cl}}[\lambda]$
algebra, where $\la$ is a free parameter, in higher spin $AdS_3$ supergravity.     

Some time ago, Kazama  and Suzuki \cite{KSNPB,KSPLB}
have found a new class of unitary ${\cal N}=2$ superconformal field
theories via coset space method.
They classified the list of Hermitian symmetric spaces and the Virasoro
central charges for the associated ${\cal N}=2$ superconformal field
theories. 
Moreover, Hull and Spence \cite{HS} studied the
description of ${\cal N}=2$ supersymmetric extension of the Kac-Moody
algebra
in ${\cal N}=2$ superspace. It turns out that the operator product
expansions between the ${\cal N}=2$ currents are nonlinear and this
fact produces exactly the same conditions in \cite{KSNPB,KSPLB}.  
Romans \cite{Romans} has found the ${\cal N}=2$ ${\cal W}_3$ algebra 
where the higher spin multiplet has spins $(2, \frac{5}{2},
\frac{5}{2}, 3)$ \footnote{We will use this notation for the spins of ${\cal
N}=2$ multiplet. The first and last one are bosonic currents while
the middle ones are fermionic. The spin contents we are dealing with for the
multiplet in this paper take the form  $(s, s+\frac{1}{2}, s+ \frac{1}{2}, s+1)$
where the spin $s$ is an integer. That is $s= 1, 2, \cdots$. The
lowest case $(1, \frac{3}{2}, \frac{3}{2}, 2)$ corresponds to the
usual ${\cal N}=2$ stress energy tensor. For $s \geq 2$, one has
higher spin currents. For the ${\cal N}=2$ ${\cal W}_{N+1}$ algebra, the highest
spin is $s_{max}=N$ and there are $N$-multiplets whose first component
spins are $s =1, 2, \cdots, N$. }. 
One of the discrete series for the central charge
matches with the central charge of KS model on ${\bf CP}^2$ coset model.
See also the work of \cite{Odake}.
By applying the ${\cal N}=2$ current algebra in \cite{HS,RASS} to the supersymmetric 
WZW conformal field theory, the explicit ${\cal N}=2$ ${\cal W}_3$
current with spins $(2, \frac{5}{2}, \frac{5}{2}, 3)$ in the above 
${\bf CP}^2$ KS model has been found
in \cite{Ahn94}.  
The free field realization was discussed in \cite{Ahn93}.
Moreover, the ${\cal N}=2$ ${\cal W}_4$ algebra was constructed in \cite{BW}
by adding one more higher spin current with spins $(3, \frac{7}{2},
\frac{7}{2},4)$ and they predicted the self-coupling constant, for the
lowest higher spin current above, which is valid for any ${\cal N}=2$ ${\cal
  W}_{N+1}$ algebra. 

In this paper, we would like to see the $AdS_3/CFT_2$ 
correspondence initiated by \cite{CHR} more detail in the context of
supersymmetric
WZW model. Contrary to the purely bosonic case where the operator
product expansion between the spin $3$ and itself does not contain the
spin $3$ current in the right hand side, the ${\cal N}=2$
supersymmetric model has the operator product expansion between the
multiplet with spins $(2, \frac{5}{2}, \frac{5}{2}, 3)$ and itself
where the right hand side contains this multiplet itself. 
For the bosonic case, it is obvious that the spin $3$ current can
occur in the $\frac{1}{(z-w)^3}$ term. However, due to the symmetry in
the operator product expansion of this current and itself, one can
also obtain the same operator product expansion by interchanging the
arguments between $z$ and $w$ and it turns out in this case, there
exists the same spin $3$ current in the above singular term but minus sign.
Then it automatically becomes zero. 
This implies that the nontrivial self-coupling in the bosonic case
occurs for the next spin $4$ where the right hand side has this spin
$4$ current in the $\frac{1}{(z-w)^4}$ singular term and we do not see
any trivial condition like as above because of the even power of this
singular term \footnote{
\label{foot}
Some time ago, the self-coupling constant for the spin $4$ current was
obtained in \cite{Hornfeck} for $W_N$ minimal model using the free
field realization.
It depends on the central charge $c$ and the $N$ explicitly. 
See also the recent paper by Gaberdiel and Gopakumar \cite{GG1} where
one can find other relevant papers.  
As far as I know, so far, there is no direct construction for this
self-coupling constant from the
operator product expansion between the $SU(N)$ 
Casimir spin $4$ operator \cite{Ahn2011} and itself. 
It would be interesting to see this feature 
although there will be lots of work to be
done for this computations.}. 

What happens if there are ${\cal N}=2$ supersymmetries in two dimensions?
One sees the presence of self-coupling even in the operator product
expansion for the lowest higher spin current.  
One simple example is the operator product expansion for the spin $2$
current and itself  that is the first component of the above multiplet 
$(2, \frac{5}{2}, \frac{5}{2}, 3)$.
One can analyze this situation as above. The spin $2$ current occurs
in the $\frac{1}{(z-w)^2}$ term in the right hand side. Due to the
even power of this singular term, there is a nontrivial spin $2$
current in this singular term. 
It is easy to see that there are no other self-coupling terms 
except this spin $2$-spin $2$ operator product expansion.
Note that there exists usual spin $2$ stress energy tensor and in
${\cal N}=2$ KS ${\bf CP}^N$ model, the higher spin current contains
other spin $2$ current which is contained in the above multiplet. 
Of course, the spin $3$-spin $3$ operator product expansion
can generate the self-coupling constant term but 
as we explained in the previous paragraph, this does not give us the
self-interacting term. 
Fortunately, it is known, in \cite{BW}, that the self-coupling
constant for the spin $2$ current in ${\cal N}=2$ ${\cal W}_{N+1}$
algebra depends on the $N$ and $k$ explicitly, as in the bosonic
case(the footnote 
\ref{foot}). 

As $N$ increases, one expects that 
there exist new primary fields in the right hand side of operator
product expansion. 
For example, the spin $3$-spin $3$ operator product expansion in
$SU(N)$ Casimir algebra leads to other spin $4$ current and its
descendant in the right hand side \cite{BBSS1,BBSS2}.
The relative coefficient functions appearing in the descendant fields for given
primary field are also fixed by conformal invariance. 

In section 2,
we rewrite the ${\cal N}=2$ current algebra in terms of the currents
living in the subgroup $H$ and the currents living in the coset
$\frac{G}{H}$ separately.
The constraints for the currents are rewritten similarly. The Sugawara
stress energy tensor is given in terms of the currents linearly or quadratically.  
For $N=2$, we describe the lowest higher spin current with explicit
group index contraction and the corresponding self-coupling constant
is given in terms of the central charge or the level.
For $N=4$, the most of the material is new. 
We present also the lowest higher spin current in terms of composite
Kac-Moody currents and explain the overall normalization constant which depends
on either the level $k$ or the central charge. We also observe the presence
of a new primary current with spins $(3, \frac{7}{2}, \frac{7}{2}, 4)$
whose structure can be described from the conformal invariance.  
For the general $N$, we notice that the self-coupling constant for
arbitrary $N$ was determined form unitarity arguments in \cite{BW}.   

In section 3, we take the large $(N,k)$ limit of the operator product
expansion between the lowest higher spin current and itself in the
context of ${\cal N}=2$ ${\cal W}_{N+1}$ algebra.

In section 4, based on the section 3, we compare the result of section
3 with the classical ${\cal N}=2$ ${\cal W}_{\infty}^{\rm{cl}}[\la]$
algebra developed in \cite{HP}.
We will present the three bosonic operator product expansions.
At linear order, one sees an agreement between the boundary and bulk theories.

In section 5, We summarize what we have found in this paper and make
some comments on the future directions.

In the Appendix, we describe  some details discussed 
in the sections $2$, $3$, $4$.

There exist some related works in \cite{Vasiliev}-\cite{GGHR}, along
the line of \cite{GG}.
       
\section{The ${\cal N}=2$ current algebra, Kazama-Suzuki coset model
and ${\cal W}_{N+1}$ algebra}

Let us consider the hermitian symmetric space
where the complex structure is preserved \footnote{Following the
  procedure in \cite{CG}, the ${\cal N}=1$ supersymmetric coset can be
written in terms of the bosonic coset \cite{CHR,CG,HP} 
by introducing the $SO(2N)$
factor in
the numerator related to the free fermions. }
\bea
{\bf CP}^N = \frac{SU(N+1)}{SU(N) \times U(1)}.
\label{cosetcpn}
\eea
Let $G=SU(N+1)$ be an even-dimensional Lie group with complex
structure and let $H=SU(N) \times U(1)$ be an even-dimensional
subgroup. This implies that the $N$ should be even.
We introduce a complex basis for the Lie algebra in which the complex
structure is diagonal and let us label the index of the group generators  
by $A$ and $\bar{A}$ where $A=1, 2, \cdots, \frac{1}{2} \mbox{dim} \, G =
\frac{1}{2} [(N+1)^2-1]$(similarly $ \bar{A} =
\bar{1}, \bar{2}, \cdots, \overline{\frac{1}{2} \mbox{dim} \, G} =
\overline{\frac{1}{2} [(N+1)^2-1]}$).
For the Hermitian generators, one has $T_{\bar{A}} = T_A^{\dagger}$
and the structure constants appear in the standard commutation
relations $[T_A, T_B] = f_{AB}^{\;\;\;\;C} T_C, [T_A, T_{\bar{B}}] =
f_{A\bar{B}}^{\;\;\;\;C} T_C+f_{A\bar{B}}^{\;\;\;\;\bar{C}} T_{\bar{C}}$ and 
$[T_{\bar{A}}, T_{\bar{B}}] = f_{\bar{A}\bar{B}}^{\;\;\;\;\bar{C}} T_{\bar{C}}$.
In other words, the structure constants $f_{AB}^{\;\;\;\;\bar{C}}$ and 
$f_{\bar{A}\bar{B}}^{\;\;\;\;C}$ vanish.
Furthermore, there are relations, 
$\mbox{Tr} (T_A T_B)=0$, $\mbox{Tr} (T_A
T_{\bar{B}})=\delta_{A\bar{B}}$, and $\mbox{Tr} (T_{\bar{A}} T_{\bar{B}})=0$. 

Then the ${\cal N}=2$ current algebra can be described by 
the ${\cal N}=2$ currents $Q^A(Z)$ and $Q^{\bar{A}}(Z)$ with nonlinear
constraints
where $Z$ stands for ${\cal N}=2$ superspace coordinates, one real
bosonic coordinate $z$, and pair of two conjugate Grassman coordinates
$\theta, \bar{\theta}$: $Z=(z, \theta, \bar{\theta})$. 
We consider the chiral currents where they are annihilated by $D_{-}$
and $\overline{D}_{-}$ and for simplicity we use $D$ for $D_{+}$ and
$\overline{D}$
for $\overline{D}_{+}$. We present the ${\cal N}=2$ current algebra in
the Appendix $A$. 

In order to obtain the generalization of Sugawara construction, it is
convenient to decompose the group $G$ indices into the subgroup $H$
indices and the coset $\frac{G}{H}$ indices explicitly. 
Let lower case middle roman indices $m, n, p, \cdots $, running from $1$
to $\frac{N^2}{2}$, refer to 
the Lie algebra of $H$, lower case top roman indices $a, b, c,
\cdots$, running from $\frac{N^2}{2}+1$ to 
$\frac{1}{2} \left[ (N+1)^2-1 \right]$,
refer to the remaining Lie algebra generators corresponding to the
coset $\frac{G}{H}$. The complex conjugated indices $\bar{m}, \bar{n},
\bar{p}, \cdots $ and $\bar{a}, \bar{b}, \bar{c}, \cdots$ hold similarly.
That is,
\bea
m, n, p, \cdots & = & 1, 2, 3, \cdots, \frac{N^2}{2}, \qquad
a, b, c, \cdots = \frac{N^2}{2}+1, \cdots, 
\frac{1}{2} \left[ (N+1)^2-1 \right],
\nonu \\
\bar{m}, \bar{n}, \bar{p}, \cdots & = & \bar{1}, \bar{2}, \bar{3}, \cdots, 
\overline{\frac{N^2}{2}}, \qquad
\bar{a}, \bar{b}, \bar{c}, \cdots = \overline{\frac{N^2}{2}+1}, 
\cdots, 
\overline{\frac{1}{2} \left[ (N+1)^2-1 \right]}.
\label{indices}
\eea
The indices $A, B, C, \cdots$ corresponding to the group $G$ 
are grouped into $m, n, p, \cdots$ of the subgroup $H$ and
$a, b, c, \cdots$ corresponding to the coset $\frac{G}{H}$. 
For the currents $Q^A(Z)$ and $Q^{\bar{A}}(Z)$,  one uses  
$J^a(Z), J^{\bar{a}}(Z)$ that live in the coset $\frac{G}{H}$, 
and $K^{m}(Z), K^{\bar{m}}(Z)$ that live in the subgroup $H$:
\bea
Q^A(Z), Q^{\bar{A}}(Z) \rightarrow K^{m}(Z), \,\,\,
K^{\bar{m}}(Z), \,\,\, J^a(Z), \,\,\, J^{\bar{a}}(Z).
\label{newfields}
\eea

Then the original operator product expansions (\ref{OPEQQ}) can be 
reexpressed in terms of the currents (\ref{newfields}) where the subgroup index
structure and remaining index structure are manifest.  
The ten(the all possibility among four currents) 
operator product expansions between these currents  are
\bea
 K^m (Z_{1}) K^n (Z_{2})   & = & -\tzb f_{\bar{m}\bar{n}}^{\;\;\;\;\bar{p}} K^p(Z_2)
-\tzbb \frac{1}{(k+N+1)} f_{\bar{m} r}^{\;\;\;\;\bar{p}}
 f_{\bar{n} \bar{r}}^
{\;\;\;\;\bar{q}}
 K^p K^q(Z_2) +\cdots,
\nonu \\
 J^a (Z_{1}) J^b (Z_{2})   & = & 
-\tzbb \frac{1}{(k+N+1)} f_{\bar{a} m}^{\;\;\;\;\bar{c}}
 f_{\bar{b} \bar{m}}^
{\;\;\;\;\bar{d}}
 J^c J^d(Z_2)+\cdots,
\nonu \\
 K^m (Z_{1}) J^a (Z_{2})   & = & -\tzb f_{\bar{m} \bar{a}}^{\;\;\;\;\bar{b}} J^b(Z_2)
-\tzbb \frac{1}{(k+N+1)} f_{\bar{m} n}^{\;\;\;\;\bar{p}}
 f_{\bar{a} \bar{n}}^
{\;\;\;\;\bar{b}}
 K^p J^b(Z_2)+\cdots,
\nonu \\
 K^{\bar{m}} (Z_{1}) K^{\bar{n}} (Z_{2})   
& = & -\tz f_{m n}^{\;\;\;\;p} K^{\bar{p}}(Z_2)
+\tzbb \frac{1}{(k+N+1)} f_{m \bar{p}}^{\;\;\;\;q}
 f_{n p}^
{\;\;\;\;r}
 K^{\bar{q}} K^{\bar{r}}(Z_2)+\cdots,
\nonu \\
 J^{\bar{a}} (Z_{1}) J^{\bar{b}} (Z_{2})   & = & 
\tzbb \frac{1}{(k+N+1)} f_{a \bar{m}}^{\;\;\;\;c}
 f_{b m}^
{\;\;\;\;d}
 J^{\bar{c}} J^{\bar{d}}(Z_2)+\cdots,
\nonu \\
 K^{\bar{m}} (Z_{1}) J^{\bar{a}} (Z_{2})   & = & -\tz f_{m
 a}^{\;\;\;\;b} 
J^{\bar{b}}(Z_2)
+\tzbb \frac{1}{(k+N+1)} f_{m \bar{p}}^{\;\;\;\;n}
 f_{a p}^
{\;\;\;\;b}
 K^{\bar{n}} J^{\bar{b}}(Z_2)+\cdots,
\nonu \\
 K^m (Z_{1}) K^{\bar{n}} (Z_{2})  & = &
 \tzzbb \frac{1}{2} \left[(k+N+1) \delta^
{m \bar{n}}  + f_{\bar{m} p}^{\;\;\;\;\bar{q}} f_{n \bar{p}}^{\;\;\;\;q} \right]
  -\frac{1}{z_{12}} (k+N+1) \delta^{m \bar{n}} 
\nonu \\
&- &  \tzb f_{\bar{m} n}^{\;\;\;\; p} K^{\bar{p}}(Z_2)   
 - \tz f_{\bar{m} n}^{\;\;\;\; \bar{p}} K^{p}(Z_2)   
\nonu \\
&- &
    \tzbb \left[ f_{\bar{m} n}^{\;\;\;\;\bar{p}} \overline{D} K^p
 + \frac{1}{(k+N+1)} f_{\bar{m} p}^{\;\;\;\;\bar{q}} f_{n \bar{p}}^{\;\;\;\;r}
 K^q K^{\bar{r}}  \right](Z_2)
+ \cdots,
\nonu \\
 K^m (Z_{1}) J^{\bar{a}} (Z_{2})  & = &
 \tzzbb \frac{1}{2}   f_{\bar{m} n}^{\;\;\;\;\bar{p}} f_{a \bar{n}}^{\;\;\;\;p} 
     - \tzb f_{\bar{m} a}^{\;\;\;\;b} J^{\bar{b}}(Z_2)   
\nonu \\
& - &    \tzbb  \frac{1}{(k+N+1)} f_{\bar{m} p}^{\;\;\;\;\bar{q}} f_{a \bar{p}}^{\;\;\;\;b}
 K^q J^{\bar{b}}(Z_2) + \cdots,
\nonu \\ 
J^a (Z_{1}) K^{\bar{m}} (Z_{2})  & = &
 - \tz f_{\bar{a} m}^{\;\;\;\;\bar{b}} J^{b}(Z_2)   
 \nonu \\
&- &   \tzbb \left[ f_{\bar{a} m}^{\;\;\;\;\bar{b}} \overline{D} J^b
+ \frac{1}{(k+N+1)} f_{\bar{a} p}^{\;\;\;\;\bar{b}} f_{m \bar{p}}^{\;\;\;\;n}
 J^b K^{\bar{n}}  \right](Z_2) + \cdots,
\nonu \\
 J^a (Z_{1}) J^{\bar{b}} (Z_{2})  & = &
 \tzzbb \frac{1}{2} \left[(k+N+1) \delta^
{a \bar{b}}  + 
f_{\bar{a} m}^{\;\;\;\;\bar{c}} f_{b \bar{m}}^{\;\;\;\;c} 
+f_{\bar{a} c}^{\;\;\;\;\bar{m}} f_{b \bar{c}}^{\;\;\;\;m} 
\right]
  \nonu \\
&- &
\frac{1}{z_{12}} (k+N+1) \delta^{a \bar{b}} 
 - \tz f_{\bar{a} b}^{\;\;\;\;\bar{m}} K^{m}(Z_2)-
   \tzb f_{\bar{a} b}^{\;\;\;\;m}  K^{\bar{m}}(Z_2) 
\nonu \\
&- &   \tzbb \left[ f_{\bar{a} b}^{\;\;\;\;\bar{m}} \overline{D} K^m 
+ \frac{1}{(k+N+1)} \left( f_{\bar{a} m}^{\;\;\;\;\bar{c}} f_{b \bar{m}}^{\;\;\;\;d}
 J^c J^{\bar{d}} +
 f_{\bar{a} c}^{\;\;\;\;\bar{m}} f_{b \bar{c}}^{\;\;\;\;n}
 K^m K^{\bar{n}} \right)
 \right](Z_2)
\nonu \\
&+ & \cdots,
\label{basicOPE} 
\eea
where 
\footnote{There is
a mathematica package \cite{KT} on ${\cal N}=2$ superspace but one
cannot use this because in our case the right hand side of the
operator product expansion (\ref{basicOPE}) 
has nonlinear structure. This is the
limitation of this package. We thank S. Krivonos for pointing out
this. 
However, from time to time, we use this package in order to extract
the component approach and are working on \cite{Thielemans} mainly for
$N=4$ case.  }
the complex spinor covariant derivatives are given by
\bea
D
=\frac{\partial}{\partial \theta}-\frac{1}{2} \overline {\theta}
\frac{\partial}{\pa z},
\qquad 
\overline{D} =\frac{\partial}{\partial \overline{\theta}}-\frac{1}{2}
\theta \frac{\partial}{\pa z}, 
\label{DDbar}
\eea
and they satisfy the algebra
\bea
D \overline{D} + \overline{D} D \equiv \{ D, \overline{D} \}=-\frac{\partial}{\pa z}.
\label{anticomm}
\eea
We also use a simplified notation as 
\bea
\th_{12}=\th_{1}-\th_{2}, \qquad \bth_{12}=\bth_{1}-\bth_{2}, \qquad
z_{12}=z_{1}-z_{2}+\frac{1}{2}(\th_{1} \bth_{2} + \bth_{1} \th_{2}).
\label{thetathetabar}
\eea
In the first equation of (\ref{basicOPE}), the property of $f_{\bar{m}
\bar{n}}^{\;\;\;\;\bar{a}}=0=f_{ab}^{\;\;\;\;m}$ is used.
In the second equation, one also uses $f_{\bar{a}
  \bar{b}}^{\;\;\;\;\bar{m}}=0=
f_{\bar{a}\bar{b}}^{\;\;\;\;\bar{c}}=f_{m\bar{n}}^{\;\;\;\; a}$.
One obtains the third equation after one uses 
$f_{m\bar{n}}^{\;\;\;\;a}=0=f_{ab}^{\;\;\;\;m}$.
For the fourth-sixth equations, one also uses similar properties of
structure constants $
f_{m
  \bar{n}}^{\;\;\;\;\bar{a}}=0=
f_{m n}^{\;\;\;\;a}=f_{ab}^{\;\;\;\;c}$ with above vanishing structure
constants.
Also the identity $f_{a\bar{b}}^{\;\;\;\;c}=0$ is used in the
remaining equations. 
In the Appendix $B$, we present the component operator product
  expansions for (\ref{basicOPE}).

One can rewrite the constraints (\ref{constraint}), by expanding the
$G$-indices into $H$-indices and $\frac{G}{H}$-indices as above,
\bea
D K^m(Z) & = & -\frac{1}{2(k+N+1)} f_{\bar{m} n}^{\;\;\;\;\bar{p}} K^n
K^p(Z),
\nonu \\
D J^a(Z) & = &  -\frac{1}{(k+N+1)} f_{\bar{a} b}^{\;\;\;\;\bar{m}} J^b K^m(Z),
\nonu \\
\overline{D} K^{\bar{m}}
(Z)& =& -\frac{1}{2(k+N+1)} f_{m \bar{n}}^{\;\;\;\;p} K^{\bar{n}}
K^{\bar{p}}(Z),
\nonu \\
\overline{D} J^{\bar{a}}(Z) & = &  
-\frac{1}{(k+N+1)}  
f_{a \bar{b}}^{\;\;\;\;m} J^{\bar{b}} K^{\bar{m}}(Z),
\label{Constraints}
\eea
where one uses
$f_{a\bar{b}}^{\;\;\;\;\bar{c}}=0=f_{mn}^{\;\;\;\;a}=f_{ab}^{\;\;\;\;m}=
f_{\bar{a}\bar{b}}^{\;\;\;\;\bar{m}}$.
For example, the 
$\theta$ and $\bar{\theta}$ independent terms in the left hand
side can be obtained from
the corresponding quantities in the right hand side of (\ref{Constraints}). 
Note that the unconstrained ${\cal N}=2$ currents have too many 
components and we have to impose constaints 
in order to preserve the number of the independent ${\cal N}=1$
currents \cite{HS}.
As we will see the component currents explicitly, 
the unconstrained ${\cal N}=1$ affine Kac-Moody currents(or its
component currents) are relocated
into the component currents in an extended ${\cal N}=2$ superspace. 
One also obtains, from (\ref{anticomm}), 
\bea
\left[ D, \overline{D} \right] K^m(Z) &= & -\pa K^m(Z) +\frac{1}{(k+N+1)} 
f_{\bar{m} n}^{\;\;\;\;\bar{p}} \left( \overline{D} K^n K^p-
K^n \overline{D} K^p \right)(Z),
\nonu \\
\left[D, \overline{D} \right] K^{\bar{m}}(Z) &= & \pa K^{\bar{m}}(Z)
- \frac{1}{(k+N+1)} f_{m \bar{n}}^{\;\;\;\;p} \left( D K^{\bar{n}}
K^{\bar{p}}- K^{\bar{n}} D K^{\bar{p}}  \right)(Z),
\nonu \\
\left[ D, \overline{D} \right] J^a(Z) &= & -\pa J^a(Z) +\frac{2}{(k+N+1)} 
f_{\bar{a} b}^{\;\;\;\;\bar{m}} \left( \overline{D} J^b K^m-
J^b \overline{D} K^m \right)(Z),
\nonu \\
\left[D, \overline{D} \right] J^{\bar{a}}(Z) &= & \pa J^{\bar{a}}(Z)
- \frac{2}{(k+N+1)} f_{a \bar{b}}^{\;\;\;\;m} \left( D J^{\bar{b}}
K^{\bar{m}}- J^{\bar{b}} D K^{\bar{m}}  \right)(Z).
\label{ConstraintsDDB}
\eea
Also in this case, the quantities in the left hand side 
are not independent because they can be obtained from the quantities
in the right hand side(cubic or linear terms) 
where all the derivative terms can be written
in terms of quadratic currents via (\ref{Constraints}). 

The Sugawara stress energy tensor for the group $G=SU(N+1)$ 
can be written in terms of 
\bea
T_G=-\frac{1}{(k+N+1)} \left[J^a J^{\bar{a}} + K^m K^{\bar{m}}
  -\left( f_{\bar{m}
      \bar{a}}^{\;\;\;\;\bar{a}} + f_{\bar{m} \bar{n}}^{\;\;\;\;\bar{n}} \right) 
D K^{\bar{m}}-\left( f_{m \bar{a}}^{\;\;\;\; \bar{a}} 
+f_{m \bar{n}}^{\;\;\;\;\bar{n}} \right) \overline{D} K^{m}
\right].
\label{Gstress}
\eea
Note that there are linear terms in the currents as well as the
quadratic terms.
As before, by using the $H$-indices and $\frac{G}{H}$-indices 
in (\ref{Sugawara})
explicitly,
the vanishing of structure constants
$f_{\bar{a}\bar{b}}^{\;\;\;\;\bar{c}}=0=
f_{m\bar{n}}^{\;\;\;\;a}$ is used.
For the metric, 
$\delta_{a\bar{m}}=0=\delta_{m\bar{a}}$.
Similarly, the stress energy tensor for the subgroup $H=SU(N) \times U(1)$
can be written as
\bea
T_H(Z)=-\frac{1}{(k+N+1)} \left[K^m K^{\bar{m}}(Z)-f_{\bar{m}
      \bar{n}}^{\;\;\;\; \bar{n}} 
D K^{\bar{m}}(Z)-f_{m \bar{n}}^{\;\;\;\;\bar{n}} 
\overline{D} K^{m}(Z)
\right].
\label{Hstress}
\eea

Then  the stress energy tensor $T(Z)$
for the supersymmetric coset model based on ${\cal N}=2$ ${\bf CP}^N$
model is obtained, by taking the difference between (\ref{Gstress}) and (\ref{Hstress}), 
\bea
T(Z)=T_G(Z)-T_H(Z)=-\frac{1}{(k+N+1)} \left[J^a J^{\bar{a}}-f_{\bar{m}
      \bar{a}}^{\,\,\,\,\,\,\,\,\bar{a}} 
D K^{\bar{m}}-f_{m \bar{a}}^{\,\,\,\,\,\,\,\, \bar{a}} 
\overline{D} K^{m}
\right](Z).
\label{stress}
\eea
From the defining operator product expansions (\ref{basicOPE}) between the currents, 
one obtains the standard operator product expansion of ${\cal N}=2$
superconformal algebra, together with (\ref{thetathetabar}),
\bea
T (Z_{1}) T (Z_{2})=
\frac{1}{z^{2}_{12}}
\frac{c}{3}
  + \left[ \tzzbb -\tz D +\tzb \overline{D} +\tzbb \partial \right]
T (Z_2).
\label{ttope}
\eea
One can easily check that there is no singular term in the operator
product expansions between the currents 
$K^m(Z_1), K^{\bar{m}}(Z_1)$ and the stress tensor $T(Z_2)$.
The corresponding central charge depends on $N$ and $k$ as follows
\footnote{This can be described as $\frac{3 k_G}{2(k_G + \tilde{h}_G)}
  \mbox{dim} \left( \frac{G}{H} \right)$ \cite{KSNPB} where $k_G=k$, 
  $\tilde{h}_G=N+1$ is the dual Coxeter number of group $G$ and  
$\mbox{dim} \left( \frac{G}{H} \right)=2N$.}:
\bea
c(N,k) & = & c_G -c_H
\nonu \\
& = &
\frac{3}{2} ((N+1)^2-1) \left[ 1-\frac{2(N+1)}{3(k+N+1)} \right]-
\frac{3}{2} (N^2-1) \left[ 1-\frac{2 N}{3(k+1+N)} \right]-\frac{3}{2}
\nonu \\
& = & \frac{3 N k}{N+k+1}.
\label{centralcharge}
\eea
Note that the coefficients of the stress energy tensors (\ref{Gstress})
and 
(\ref{Hstress}) are the same. This is different feature from the
coset construction for the bosonic theory where the diagonal
subalgebra exists and the coefficients of the various stress energy
tensors are different. 
Also, the level of the group $SU(N+1)$ and the level of $SU(N)$ are
same as each other. In other words, each $\frac{1}{(z-w)^2}$ term of
spin $1$-spin $1$ operator expansion has the same factor $(k+N+1)$.  
For the explicit form, see the Appendix $B$.
This shift $(k+N+1)$ rather than $k$ arises from the ${\cal N}=1$ 
supersymmetrization.

There are two requirements on the ${\cal N}=2$  current 
$W(Z)$ of generators with spins $(2, \frac{5}{2}, \frac{5}{2}, 3)$.  

1) The operator product expansions between the $H$-currents 
$K^m(Z), K^{\bar{m}}(Z)$ and the $\frac{G}{H}$-current $W(Z)$
should not 
contain any singular terms:
\bea
K^m(Z_1) W(Z_2) =0, \qquad K^{\bar{m}}(Z_1) W(Z_2)=0.
\label{KWvanishing}
\eea
In practice, one uses the component approach and due to the
constraints (\ref{Constraints}), only after some of the operator
product expansions(among $16$ operator product expansions for each
case in (\ref{KWvanishing})) are checked, the coefficient fucntions appearing in
the unknown  higher spin current $W(Z)$ are determined completely
except the overall constant.   

2) The current $W(Z)$ with vanishing $U(1)$ charge 
is a ${\cal N}=2$ primary field under the stress
energy tensor (\ref{stress})  
\bea
T (Z_{1}) W (Z_{2})=
 \left[ \tzzbb 2 -\tz D +\tzb \overline{D} +\tzbb \partial \right]
W (Z_2).
\label{TW}
\eea
Here there is no term like $\frac{1}{z_{12}}$ due to the zero $U(1)$ charge.
The coefficient $2$ in $\tzzbb$ implies the lowest spin of $W(Z)$.
We present the component results for (\ref{TW}) in the Appendix $C$.
In general, there exist $\frac{1}{z_{12}^3}$-, 
$\frac{\theta_{12} \bar{\theta}_{12}}{z_{12}^3}$-,
$\frac{\theta_{12}}{z_{12}^3}$-, and 
$\frac{\bar{\theta}_{12}}{z_{12}^3}$-terms with appropriate composite
currents.
The requirement 2) implies that these extra terms should vanish by
choosing the correct coefficient functions. 

For the ${\cal N}=2$ currents, the component currents are given by
\bea
K^m(Z) & = & 
K^m(z)+ \theta \,\, D K^m(z) + \bar{\theta} \,\, 
\overline{D} K^m(z)+ \theta \bar{\theta}\,\, (-1) \frac{1}{2} [ D, \overline{D} ] K^m(z),
\nonu \\ 
K^{\bar{m}}(Z) & = & 
K^{\bar{m}}(z)+ \theta \,\,  D K^{\bar{m}}(z) +\bar{\theta} \,\,
\overline{D} K^{\bar{m}}(z) +\theta \bar{\theta} \,\,  (-1) \frac{1}{2} [
D, \overline{D} ] K^{\bar{m}}(z),
\nonu \\
J^a(Z) & = & J^a(z) +\theta \,\, D J^a(z) +\bar{\theta} \,\, \overline{D} J^a(z)+ 
\theta \bar{\theta} \,\, (-1) \frac{1}{2} [ D, \overline{D} ] J^a(z), 
\nonu \\ 
J^{\bar{a}}(Z) & = & 
J^{\bar{a}}(z) +\theta \,\, D J^{\bar{a}}(z) +\bar{\theta}  \,\,
\overline{D} J^{\bar{a}}(z) +\theta \bar{\theta} \,\, (-1) \frac{1}{2} [
D, \overline{D} ] J^{\bar{a}}(z).
\label{components}
\eea
Due to the constraints (\ref{Constraints}) and (\ref{ConstraintsDDB}),
the $\theta$- and $\theta \bar{\theta}$ components of $K^m(Z)$ and
$J^a(Z)$ are not independent but they can be written in terms of 
other independent terms. Similarly, 
the $\bar{\theta}$- and $\theta \bar{\theta}$ components of $K^{\bar{m}}(Z)$ and
$J^{\bar{a}}(Z)$ can be written in terms of 
other independent terms.

Let us emphasize how one applies the above two conditions 1) and 2). 
For the general ${\cal N}=2$ ${\cal W}_{N+1}$ algebra, we use them
in ${\cal N}=2$ superspace but for fixed $N=4$ case, we use the
package \cite{Thielemans} where the component result is necessary to
obtain the operator product expansions. Therefore, one should apply
the two conditions in the component approach. The component result for
(\ref{TW})
is summarized in the Appendix $C$. For the regularity condition,  given
in 1), among $16$ operator product expansions for each case, 
only the half of them are independent from the arguments in
(\ref{components}). Once we have checked the regularity condition for
the independent components, then the condition 1) satisfies
automatically, by construction.  We do not need to check the other
remaining half of the equations. 
 
For the stress energy tensor \footnote{Our notation corresponds to the
  one in \cite{Romans} as follows:
$T(z) \leftrightarrow J_{ro}(z), D T(z) \leftrightarrow G_{ro}^{+}, \overline{D}
  T(z) \leftrightarrow G_{ro}^{-}$, and $-\frac{1}{2} [D, \overline{D}]
  T(z) \leftrightarrow T_{ro}(z)$.}, one has  
\bea
T(Z) & = & 
T(z)+ \theta \,\, D T(z) + \bar{\theta} \,\, 
\overline{D} T(z)+ \theta \bar{\theta}\,\, (-1) \frac{1}{2} [ D, \overline{D} ] T(z),
\label{compstress}  
\eea
where the component fields can be obtained from (\ref{stress}) by
using the covariant derivatives (\ref{DDbar}) with (\ref{anticomm})
and putting the $\theta, \bar{\theta}$'s to vanish.
$T(z)$ is a $U(1)$ current of spin $1$, $D T(z)$ and $\overline{D} T(z)$
are fermionic currents of spin $\frac{3}{2}$ and $- 
\frac{1}{2} [ D, \overline{D} ] T(z)$ is the stress energy tensor of
spin $2$.

In next subsections, we will construct the higher spin currents explicitly.
Starting with $N=2$ case, one considers the $N=4$ case and 
wants to generalize for arbitrary $N$ which corresponds to 
${\cal N}=2$ ${\cal W}_{N+1}$ algebra.  

\subsection{The $N=2$ case: ${\bf CP}^2 (= \frac{SU(3)}{SU(2) 
\times U(1)})$ coset model}

The ${\cal N}=2$ ${\cal W}_3$ algebra has one additional extra higher
spin current
with spins $(2, \frac{5}{2}, \frac{5}{2}, 3)$, as well as the ${\cal
  N}=2$
superconformal algebra (\ref{ttope}), where
one can write down the following component currents explicitly \footnote{
Similarly, our fields correspond to the
ones in \cite{Romans} as follows:
$W(z) \leftrightarrow V_{ro}(z), D W(z) \leftrightarrow U_{ro}^{+}, \overline{D}
  W(z) \leftrightarrow U_{ro}^{-}$, and $-\frac{1}{2} [D, \overline{D}]
  W(z) \leftrightarrow W_{ro}(z)$.}
\bea
W(Z) & = & 
W(z)+ \theta \,\, D W(z) + \bar{\theta} \,\, 
\overline{D} W(z)+ \theta \bar{\theta}\,\, (-1) \frac{1}{2} [ D, \overline{D} ] W(z).
\label{n2W}
\eea
In this case, the number of independent WZW currents is $8$ and it is
not so complicated to write down all the possible terms for the
current (\ref{n2W}). 
However, the explicit form for this current in \cite{Ahn94} is not
convenient to generalize to the arbitrary ${\cal N}=2$ ${\cal
W}_{N+1}$ algebra because there are no any contractions between the
$SU(3)$ group indices. 

For given result for the expression of (\ref{n2W}) in \cite{Ahn94}, 
one can think of the equivalent expression  as follows \footnote{There
are $48$ nonzero structure constants.}: 
\bea
W(Z) & = &
a_1 \, f_{\bar{p} a}^{\;\;\;\;b} f_{p m}^{\;\;\;\;n} 
J^a J^{\bar{b}} K^m K^{\bar{n}}(Z) +
a_2 \, f_{p a}^{\;\;\;\;b} f_{\bar{p}m}^{\;\;\;\;n} 
 J^a J^{\bar{b}} K^m K^{\bar{n}}(Z)+
a_3 \, J^a J^b J^{\bar{a}} J^{\bar{b}}(Z) \nonu \\
& + &
 a_4 \,
f_{\bar{m} a}^{\;\;\;\;b}  J^{a} J^{\bar{b}} D K^{\bar{m}}(Z) 
 +  a_5 \, f_{m a}^{\;\;\;\;b}  J^{a} J^{\bar{b}} \overline{D} K^{m}(Z)
+a_6 \, f_{\bar{m} \bar{n}}^{\;\;\;\;\bar{n}}  
D K^{\bar{m}} J^a J^{\bar{a}}(Z)  
\nonu \\
& + &  a_7 \, f_{m \bar{n}}^{\;\;\;\;\bar{n}}  
\overline{D} K^{m} J^a J^{\bar{a}}(Z) +a_{8} \, \overline{D} J^a D
J^{\bar{a}}(Z) + a_{9} \, \overline{D} K^m D
K^{\bar{m}}(Z) \nonu \\
& + & 
a_{10} \, J^a \pa J^{\bar{a}}(Z) + a_{11} \, \pa J^a J^{\bar{a}}(Z)
 + 
a_{12} \, K^m \pa K^{\bar{m}}(Z) 
 +   a_{13} \, \pa K^m K^{\bar{m}}(Z)
\nonu \\
& + & a_{14} \, J^a [D, \overline{D}] J^{\bar{a}}(Z) +
a_{15}  \, [D, \overline{D}] J^a J^{\bar{a}}(Z)
+a_{16}  \, K^m  [D, \overline{D}] K^{\bar{m}}(Z) 
\nonu \\
& + & a_{17} \, [D, \overline{D}] K^m K^{\bar{m}}(Z)
+a_{18} \, D J^a \overline{D} J^{\bar{a}}(Z)+
a_{19} \, D K^m \overline{D} K^{\bar{m}}(Z)
\nonu \\
&  + & a_{20} \, f_{m\bar{n}}^{\;\;\;\;\bar{n}} \pa \overline{D} K^{m}(Z)
+ a_{21} \, f_{\bar{m} \bar{n}}^{\;\;\;\;\bar{n}} \pa D K^{\bar{m}}(Z)
 +  a_{22} \, f_{m \bar{p}}^{\;\;\;\;\bar{p}} 
f_{n\bar{q}}^{\;\;\;\;\bar{q}}
\overline{D} K^m \overline{D}
K^{n}(Z) 
\nonu \\
& + &  a_{23} \, f_{\bar{m} \bar{p}}^{\;\;\;\;\bar{p}} 
f_{\bar{n} \bar{q}}^{\;\;\;\;\bar{q}}
D K^{\bar{m}} D
K^{\bar{n}}(Z) +  
   a_{24} \, f_{m \bar{p}}^{\;\;\;\;\bar{p}} 
f_{\bar{n} \bar{q}}^{\;\;\;\;\bar{q}}
\overline{D} K^{m} D
K^{\bar{n}}(Z),
\label{n2terms}
\eea
where all the coefficient functions are present in the Appendix (\ref{coeffn2one}).
This explicit structure (\ref{n2terms}) was obtained from the two conditions
(\ref{KWvanishing}) 
and (\ref{TW}).
We also present the operator product expansion at the linearized level
in the Appendix $D$ where the right hand side contains 
the central charge
\bea
c_{N=2} =\frac{6k}{k+3},
\label{n2central}
\eea
and the self-coupling constant
\bea
\alpha_{N=2}^2=
\frac{27(2-k)^2(1+k)^2}{(-1+k)(5+k)(3+2k)(-3+5k)}
= -\frac{(3+c)^2 (-12+5 c)^2}{2 (-15+c) (-1+c) (6+c) (-3+2 c)},
\label{const1}
\eea
where we replace the level $k$ with the central charge $c$ (\ref{n2central}).
Compared to the pure bosonic case(for example, the operator product
expansion between the spin $3$ current and itself in terms of WZW
currents), as in introduction,
the operator product expansion of $W(Z)$ and itself in ${\cal N}=2$
superspace or in the component approach has a self-coupling constant
term in the right hand side.
For the bosonic spin $3$ case, there is no spin $3$ current that will
appear in the $\frac{1}{(z-w)^3}$ term  in the
right hand side of the operator product expansion. 
One can easily see this observation by considering the operator
product expansion with reversing the arguments and realizing that
there will be an inconsistency in the operator product expansion.
However, this is not true for the spin $4$ case. In general, the
operator product expansion between the spin $4$ current and itself(in
terms of WZW currents) in
$W_N$ algebra generates the spin $4$ current in the right hand side.           
It would be interesting to find out the self-coupling constant for the
spin $4$ current in the bosonic case. 

We present the operator product expansion in (\ref{open2linear}) at
linearized level. Note that the coefficient functions in the right
hand side are characterized by the central charge $c$ and
self-coupling constant $\alpha$. One sees that the $\alpha$ dependence
appears in the current $W(Z_2)$ and its descendant fields and the
functions of central charge appear in the other remaining fields. 

\subsection{The $N=4$ case:  ${\bf CP}^4 
(=\frac{SU(5)}{SU(4) \times U(1)})$ coset model}
\label{n4}

Let us recall that the field contents of ${\cal N}=2$ ${\cal W}_5$
algebra are given by the stress energy tensor with spins $(1,
\frac{3}{2}, \frac{3}{2}, 2)$ and higher spin currents with spins $(2,
\frac{5}{2}, \frac{5}{2}, 3),
(3, \frac{7}{2}, \frac{7}{2}, 4)$, and $(4, \frac{9}{2}, \frac{9}{2},
5)$ \footnote{If one considers ${\cal N}=2$ ${\cal W}_4$ algebra, then
the coset can be described as ${\bf CP}^3 =\frac{SU(4)}{SU(3) \times
  U(1)} =\frac{SU(4) \times U(1)}{SU(3) \times U(1) \times
  U(1)}=\frac{U(4)}{U(3) \times U(1)}$ by
introducing the extra $U(1)$'s in order to have even-dimensional groups
$G$ and $H$  from (\ref{cosetcpn}). 
In principle, one can find the corresponding ${\cal N}=2$
current algebra with $U(4)$ group in the complex basis. This should
correspond to the work of \cite{BW}.}. 
Then how one can determine these currents in terms of the fundamental
currents which live in the supersymmetric WZW model?
As before, the stress energy tensor is given by (\ref{stress}).
It is nontrivial to find the extra symmetry currents in 
the generalization of Sugawara construction(so called Casimir
construction) that includes the higher spin generators. 
Compared to the previous case where there exist only 
$8$ independent fields, there exist $24$ independent
fundamental WZW currents.
One way to write down the lowest higher spin
current with spins $(2, \frac{5}{2}, \frac{5}{2}, 3)$ is
to take into account of all the possible terms(quartic terms, cubic
terms and quadratic terms and linear terms in the WZW currents (\ref{newfields})).   
The other way is to take the $N=2$ case (\ref{n2terms}) with arbitrary
coefficient functions and apply the two conditions (\ref{KWvanishing})
and 
(\ref{TW}) but did not come out properly.
By brute force, one should add other possible terms coming from 
\bea
TT(Z),\,\,\, \pa T(Z),\,\,\, [D, \overline{D}]T(Z),\,\,\, 
T_H T_H(Z), \,\,\,\pa T_H(Z), \,\,\,[D,
\overline{D}] T_H(Z), \,\,\, T T_H(Z),
\label{spin2contents}
\eea
where $T(Z)$ is given by (\ref{stress}) and $T_H(Z)$ is given by
(\ref{Hstress}).
In other words, by looking at the explicit expressions
(\ref{spin2contents}), collecting the independent terms 
and adding these into the expression (\ref{n2terms}).
Finally, it turns out that the correct higher spin current with spins
$(2, \frac{5}{2}, \frac{5}{2}, 3)$, satisfying the  two conditions (\ref{KWvanishing})
and (\ref{TW}), takes the form
\bea
W(Z) &=& 
b_1 \, f_{\bar{c} a}^{\;\;\;\;\bar{m}} f_{c\bar{b}}^{\;\;\;\;n} 
J^a J^{\bar{b}} K^m K^{\bar{n}}(Z) +
b_2 \, f_{\bar{c} a}^{\;\;\;\;n} f_{c\bar{b}}^{\;\;\;\;\bar{m}} 
 J^a J^{\bar{b}} K^m K^{\bar{n}}(Z)+
b_3 \, J^a J^b J^{\bar{a}} J^{\bar{b}}(Z) \nonu \\
& + &
b_4 \, f_{a\bar{m}}^{\;\;\;\;b} J^a K^{\bar{m}} D J^{\bar{b}}(Z) 
+ b_5  \, f_{m a}^{\;\;\;\;b} K^m \overline{D} J^{a} J^{\bar{b}}(Z) 
+ b_6 \,
f_{\bar{m} a}^{\;\;\;\;b}  J^{a} J^{\bar{b}} D K^{\bar{m}}(Z) \nonu \\
& + & b_7 \, f_{m a}^{\;\;\;\;b}  J^{a} J^{\bar{b}} \overline{D} K^{m}(Z)
+b_8 \, f_{m n}^{\;\;\;\;p}  \overline{D} K^{m} K^n K^{\bar{p}}(Z)  
+  b_9 \, \overline{D} J^a D J^{\bar{a}}(Z) 
\nonu \\
& + & b_{10} \,  \overline{D} K^m D
K^{\bar{m}}(Z)
+b_{11} \, J^a \pa J^{\bar{a}}(Z) + b_{12} \, \pa J^a J^{\bar{a}}(Z)
 + 
b_{13} \, K^m \pa K^{\bar{m}}(Z) 
\nonu \\
& + &  b_{14} \, \pa K^m K^{\bar{m}}(Z)
 + b_{15} \, J^a [D, \overline{D}] J^{\bar{a}}(Z) +
b_{16}  \, [D, \overline{D}] J^a J^{\bar{a}}(Z)
+b_{17}  \, K^m  [D, \overline{D}] K^{\bar{m}}(Z) 
\nonu \\
& + & b_{18} \, [D, \overline{D}] K^m K^{\bar{m}}(Z)
+b_{19} \, D K^m \overline{D} K^{\bar{m}}(Z)
 + b_{20} \, f_{m\bar{n}}^{\;\;\;\;\bar{n}} \pa \overline{D} K^{m}(Z)
+ b_{21} \, f_{\bar{m} \bar{n}}^{\;\;\;\;\bar{n}} \pa D K^{\bar{m}}(Z)
\nonu \\ 
& + & b_{22} \, f_{\bar{m} \bar{b}}^{\;\;\;\;\bar{b}} J^a J^{\bar{a}} D
K^{\bar{m}}(Z)
+ b_{23}  \, f_{m \bar{b}}^{\;\;\;\;\bar{b}} J^a J^{\bar{a}} \overline{D}
K^{m}(Z)
+ b_{24} \, f_{\bar{m}\bar{a}}^{\;\;\;\;\bar{a}}
  f_{\bar{n}\bar{b}}^{\;\;\;\;\bar{b}} D K^{\bar{m}} D K^{\bar{n}}(Z)
\nonu \\
& + & b_{25} \, f_{\bar{m}\bar{a}}^{\;\;\;\;\bar{a}}
  f_{n \bar{b}}^{\;\;\;\;\bar{b}} D K^{\bar{m}} \overline{D} K^{n}(Z)
+b_{26} \, f_{m\bar{a}}^{\;\;\;\;\bar{a}}
  f_{n \bar{b}}^{\;\;\;\;\bar{b}} \overline{D} K^{m}
  \overline{D} K^{n}(Z)
 +b_{27} \,
f_{\bar{m} a}^{\;\;\;\;a } \pa D K^{\bar{m}}(Z),
\label{superspin2}
\eea
where all the coefficient functions are given in the Appendix $F$
explicitly \footnote{Totally, there are $249$ independent terms if we
expand out the
  structure constants(the number of nonzero structure constants is
  $492$ from the discussion of the Appendix $E$)
and the metric. }.
This is an ${\cal N}=2$ current and one can read off the corresponding
component currents with spins $(2, \frac{5}{2}, \frac{5}{2}, 3)$.
The spin $2$ current $W(z)$ in (\ref{n2W}) can be obtained by putting all the
$\theta$ and $\bar{\theta}$ dependence in the right hand side of
(\ref{superspin2}) 
to zero. For the spin $\frac{5}{2}$ currents $D W(z)$ and
$\overline{D} W(z)$ can be obtained also by taking the supercovariant
derivatives $D$ and $\overline{D}$ into the right hand side of
(\ref{superspin2})
and then putting the $\theta$ and $\bar{\theta}$ to zero at the final expression.
For the spin $3$ current $-\frac{1}{2} [D, \overline{D}] W (z)$, 
one can do similar analysis.
Due to the constraints (\ref{Constraints}),
there are several ways to write down these component currents \footnote{
Also note that the previous spin $2$ current (\ref{n2terms}) can be
written in terms of (\ref{superspin2}) with the coefficients in the
Appendix (\ref{coeffn2two}).}.

On the other hands, one can make the explicit operator product
expansion between $T(Z_1)$ (\ref{stress}) and $W(Z_2)$
(\ref{superspin2}) by using the defining equations (\ref{basicOPE}).
Since it should satisfy the primary condition (\ref{TW}), one can read
off the above component currents straightforwardly by looking at the
singular terms in the operator product expansion.
We list them in the Appendix $C$.
How to determine the spin $3$ current, for example?
First, we determine the spin $\frac{5}{2}$ current $\overline{D} W(z)$ 
by using the seventh
equation of (\ref{TWcomp}) and reading off the $\frac{1}{(z-w)}$
terms where one uses $\overline{D} T(z)$ from (\ref{compstress}).
Next, by using  the fifth equation of (\ref{TWcomp}) 
with spin $\frac{3}{2}$ current $D T(z)$ (\ref{compstress}) and collecting the
$\frac{1}{(z-w)}$ terms, one sees the spin $3$ current
$-\frac{1}{2}[D, \overline{D}] W(w)$ and the descendant field $\pa
W(w)$. During this computation, one uses the constraint equations 
(\ref{Constraints}) all
the time.
In this way, one obtains all the component fields.
For example, the field $D W(w)$ can be obtained from the fourth
equation of (\ref{TWcomp}). 

Let us focus on the $\frac{1}{(z-w)^4}$ terms in the operator product
expansion of $W(z) W(w)$  where the spin $2$ current $W(z)$ is the first
component of $W(Z)$ in (\ref{n2W}) that has the form in
(\ref{superspin2}) together with (\ref{coeffn5}).
One determines the overall constant $A(k)$ 
\bea
A(k)^2 & = &
-\frac{\left(25 \sqrt{3}+135 i \sqrt{5}-23 \sqrt{3} 
k+15 i \sqrt{5} k-8 \sqrt{3} k^2\right)^2}{8 (-1+k) (5+k)^4 (9+k) (5+2 k) (-5+11 k)}
\nonu \\
 & = &
\frac{(-12+c)^4 \left(-60 \sqrt{3}-324 i \sqrt{5}+33 \sqrt{3} c+39 
i \sqrt{5} c+\sqrt{3} c^2-i \sqrt{5} c^2\right)^2}{207360000 (-27+c)
(-2+c) (-1+c) 
(12+c)},
\label{Ak}
\eea
by requiring  that the $\frac{1}{(z-w)^4}$ term should be equal to
$\frac{c}{2}$
where the central charge is
\bea
c_{N=4} = \frac{12 k}{k+5}.
\label{centraln4}
\eea

Let us consider the $\frac{1}{(z-w)^2}$ terms 
 in the operator product
expansion of $W(z) W(w)$. In general, there are three different field
 contents,
$W(w), [D, \overline{D}]T(w)$ and $T T(w)$.
The easiest way to determine the self-coupling constant appearing in
 the coefficient function in front of $W(w)$(in the right hand side of
 this operator product expansion $W(z) W(w)$) is to focus on 
any quartic term which does not appear in the fields 
$[D, \overline{D}]T(w)$ and $T T(w)$. For example, 
let us consider the $K^1 K^5 K^{\bar{3}} K^{\bar{7}}(w)$ in the
 $\frac{1}{(z-w)^2}$
term. Definitely, this quartic term does not appear in the $[D,
 \overline{D}]T(w)$ and 
$T T(w)$.
It turns out that the self-coupling constant is given in terms of
 either
the level $k$ or the central charge $c$ (\ref{centraln4}):  
\bea
\alpha_{N=4}^2  = 
\frac{25 (-4+k)^2 (1+k)^2}{(-1+k) (9+k) (5+2 k) (-5+11 k)}
 =  \frac{(3+c)^2 (-16+3 c)^2}{2 (27-c) (-2+c) (-1+c) (12+c)}.
\label{const2}
\eea
Compared to the previous one for $N=2$ (\ref{const1}), this is
different from (\ref{const1}). 
It seems that the factors $(3+c)^2$ and $(-1+c)$ are common and they
appear as $N$-independent factors but other factors should behave
as $N$-dependent factors. 
Therefore, one should consider the most general self-coupling constant
which will depend on $N$ 
when one describes the Kazama-Suzuki model for the general $N$.

Let us consider the $\frac{1}{(z-w)^2}$ term in the operator product
expansion of the spin $2$ field and the spin $3$ field, 
$W(z) (-1)\frac{1}{2} [D, \overline{D}] W(w)$.
We do not present the spin $3$ current here because the full
expression for this is rather complicated.
In general, there exist the seven different spin 
$3$ fields in this singular term:
\bea
[D, \overline{D}]W(w), \,\, T
W(w), \,\, \pa [D, \overline{D}] T(w), \,\, T [D, \overline{D}]T(w),
\,\, T T
T(w), \,\, \overline{D} T D T(w), \,\, \pa^2 T(w).
\label{spin3comp}
\eea  
As in bosonic case \footnote{Recall that for the bosonic case, the spin $4$ current appearing 
in the operator
product expansion between the spin $3$ current and itself
vanishes for $N=3$ and one of the levels being $1$ in the $W_N$ coset minimal model.
However, for general $N$, the spin $4$ current occurs naturally.} 
\cite{Ahn2011}, one expects that there should be the extra higher
spin fields because the field contents of ${\cal N}=2$
${\cal W}_{5}$ algebra are given by the multiplet $W(Z)$ with spins 
$(2,\frac{5}{2}, \frac{5}{2},3)$, 
the multiplet $V(Z)$ 
with spins $(3, \frac{7}{2}, \frac{7}{2},4)$, and the multiplet $X(Z)$ with spins 
$(4, \frac{9}{2},\frac{9}{2},5)$.
As before, one has the following component currents for this new
primary current with spins $(3, \frac{7}{2}, \frac{7}{2}, 4)$, for example,
\bea
V(Z) & = & 
V(z)+ \theta \,\, D V(z) + \bar{\theta} \,\, 
\overline{D} V(z)+ \theta \bar{\theta}\,\, (-1) \frac{1}{2} [ D, \overline{D} ] V(z).
\label{Vexpress}
\eea

Although it is rather involved procedure to extract the exact form for
the new primary field explicitly, one can check the existence of this
field by looking at the particular term in the corresponding
$\frac{1}{(z-w)^2}$ term.
It turns out that, in ${\cal N}=2$ superspace,
one should add the following extra singular terms in the operator
product expansion $W(Z_1) W(Z_2)$, at the linearized level,
compared to the ${\cal N}=2$ ${\cal
W}_3$ algebra described in previous subsection,
\bea
\frac{\theta_{12} \bar{\theta}_{12}}{z_{12}^2} \,\, 3 V(Z_2) +
\frac{\bar{\theta}_{12}}{z_{12}} \,\, \overline{D} V(Z_2) -
\frac{\theta_{12} }{z_{12}} \,\, D V(Z_2) +
\frac{\theta_{12} \bar{\theta}_{12}}{z_{12}} \,\, 2 \pa V(Z_2).
\label{extra}
\eea
It is easy to see that all the relative coefficients can be fixed by
using the conformal invariance, for given normalized factor $3$ in
front of $V(Z_2)$ in (\ref{extra}).
In the operator product expansion of $\phi_m(z) \phi_n(w) \sim \phi_p +
\mbox{descendant}$, the relative coefficient function, for given
primary field, can be
determined by conformal invariance \cite{BPZ,Dotsenko}. 

For example, let us put the arbitrary coefficients $c_1$ and $c_2$ and
$c_3$ in the second, third and fourth term respectively.
Then the operator product expansion of $W(z) [D, \overline{D}]W(w)$(
see also (\ref{ope346})) 
can be written in terms of 
\bea
\frac{1}{(z-w)^2} (-1) 6 V(w) +\frac{1}{(z-w)} 
\left[ (c_1-c_2 -2 c_3) \pa V + \cdots \right](w) +\cdots.
\label{ope233}
\eea
The descendant field of $V(w)$, at $\frac{1}{(z-w)}$ term, is given by $\pa V(w)$.
We do not write down other terms which are not relevant to our consideration. 
One uses the formula for the relative coefficient 
is given by 
\bea
\frac{h_p +h_m -h_n}{2 h_p}= \frac{3 + 2 -3}{2 \times 3} =\frac{1}{3},
\label{formula}
\eea
where the field $\phi_m$ plays the role of $W(z)$ which has a
conformal dimension $2$, the field $\phi_n$ corresponds to 
$[D, \overline{D}]W(w)$ with spin $3$ and the field
$\phi_p$ corresponds to $V(w)$ with spin $3$.
Therefore, the coefficient  of $\pa V(w)$ should be equal to $-2$ for
given the coefficient $-6$ in front of $V(w)$ in (\ref{ope233}).
That is, $\frac{-2}{-6}= \frac{1}{3}$.
Then one has
\bea
c_1-c_2 -2 c_3 =-2.
\label{1}
\eea
Similarly, the operator product expansion $D W(z) \overline{D}
W(w)$(see also (\ref{ope351}) and (\ref{ope352}))
contains
\bea
\frac{1}{(z-w)^2} 3 V(w) +\frac{1}{(z-w)} 
\left[ \frac{1}{2}(c_2+ 2 c_3) \pa V + \cdots \right](w) +\cdots.
\label{opeother}
\eea
By counting the conformal dimensions and using the above formula, one
gets
\bea
\frac{h_p +h_m -h_n}{2 h_p} =
\frac{3+\frac{5}{2}-\frac{5}{2}}{2 \times 3}=\frac{1}{2}.
\nonu
\eea
The coefficient of  $\pa V(w)$ in (\ref{opeother}) 
should be equal to $\frac{3}{2}$.
Therefore, the following relation holds
\bea
 \frac{1}{2}(c_2+ 2 c_3) = \frac{3}{2}.
\label{2}
\eea
Due to the structure of the formula, 
as long as $h_m=h_n$, the relative coefficient becomes 
$\frac{1}{2}$ \cite{BPZ}.
Finally, the operator product expansion 
$D W(z) [D, \overline{D}] W(w)$ has the singular term
\bea 
\frac{1}{(z-w)^2} (-6 +c_2) D V(w) +\frac{1}{(z-w)} 
\left[ (-c_2- 2 c_3) \pa D V + \cdots \right](w) +\cdots.
\label{opeopeother}
\eea
Again, in this case, one has 
\bea
\frac{h_p +h_m -h_n}{2 h_p} = \frac{\frac{7}{2}+ \frac{5}{2}-3}{2
  \times 
\frac{7}{2}} = \frac{3}{7}.
\nonu
\eea
The coefficient of  $\pa D V(w)$ in (\ref{opeopeother}) 
should be equal to $\frac{3}{7}(-6+c_2)$.
The final equation satisfies
\bea
 (-c_2- 2 c_3) = \frac{3}{7}(-6+c_2).
\label{3}
\eea
By combing these three equations (\ref{1}), (\ref{2}) and (\ref{3}), 
and solving them, then there exists a
unique solution and one has $c_1=1, c_2 =-1$, and $c_3=2$.

Due to the field contents for ${\cal N}=2$ ${\cal W}_5$ algebra,
there are other five operator product expansions in ${\cal N}=2$
superspace.
In principle, one obtains them by taking the operator product
expansions once all the primary fields are determined and expressed in
terms of WZW currents, although the computations will be rather complicated. 

\subsection{The general $N$ case:  ${\bf CP}^N (=\frac{SU(N+1)}{SU(N)
    \times U(1)})$ 
coset model}

The self-coupling constant of the current with spins $(2, \frac{5}{2},
\frac{5}{2},3)$ for any ${\cal N}=2$ ${\cal W}_{N+1}$ algebra
is determined from the unitarity arguments in \cite{BW}
\bea
\alpha(N,k)^2 & = &
\frac{3 (1+k)^2 (k-N)^2 (1+N)^2}{(-1+k) (-1+N) (1+2 k+N) (1+k+2 N)
  (-1-N+k (-1+3 N))}
\nonu \\
& = & \frac{(3+c)^2 
\left(c+2 c N-3 N^2\right)^2}{(-1+c) (3-c+6 N) (-1+N) (c+3 N) (-3 N+c (2+N))},
\label{selfcoupling}
\eea
where
the central charge is given by (\ref{centralcharge}).
Note that the previous constants (\ref{const1}) and (\ref{const2})
can be read off from this general expression (\ref{selfcoupling}) by
substituting $N=2$ and $N=4$ respectively.

This behavior is also observed in the work of \cite{HP} by considering the
singular terms $\frac{1}{(z-w)^4}$ and $\frac{1}{(z-w)^2}$ with spin
$2$ current in the KS model simultaneously because the normalization
in the highest
singular term is different from each other.  
They computed the operator product expansions between the
spin $2$ field and itself, by following \cite{Ito} in the context of
free field realization, for $N=2,3,4,5$ cases 
and obtained by extrapolating these results. 

Now one can write down the operator product expansion between $W(Z_1)$
and $W(Z_2)$ as follows:
\bea
&& W(Z_1) W(Z_2)  = \nonu \\
&& \frac{1}{z_{12}^4} \,\, \frac{c}{2} +\frac{\theta_{12}
  \bar{\theta}_{12}}{z_{12}^4}
\,\, 3 T(Z_2) +\frac{\bar{\theta}_{12}}{z_{12}^3} \,\, 3 \overline{D} T(Z_2)
-\frac{\theta_{12}}{z_{12}^3} \,\, 3 D T(Z_2) +\frac{\theta_{12} 
\bar{\theta}_{12}}{z_{12}^3} \,\, 3 \pa T(Z_2)
\nonu \\
&& + \frac{1}{z_{12}^2} \left[ 2 \alpha W +\frac{c}{(-1+c)} [D,
  \overline{D}] T \right](Z_2) +\frac{\bar{\theta}_{12}}{z_{12}^2}
 \left[ \alpha \overline{D} W +\frac{(-3+2c)}{(-1+c)} 
  \pa \overline{D} T \right](Z_2)
\nonu \\
&& +  \frac{\theta_{12}}{z_{12}^2}
 \left[ \alpha D W -\frac{(-3+2c)}{(-1+c)} 
  \pa D T \right](Z_2) 
 + 
\frac{\theta_{12} \bar{\theta}_{12}}{z_{12}^2} \left[ 
\frac{3(-8+c)}{2(-12+5c)} \alpha [D, \overline{D} ] W \right. \nonu \\ 
&& +  \left. 
\frac{9c(-12+5c)}{4(-1+c)(6+c)(-3+2c)} \pa [D, \overline{D}] T +
\frac{3(18-15c+2c^2+2c^3)}{2(-1+c)(6+c)(-3+2c)} \pa^2 T + 3 V \right](Z_2)
\nonu \\
&& +  \frac{1}{z_{12}} \left[  
\alpha \pa  W -\frac{c}{2(-1+c)} \pa [D,
  \overline{D}] T \right](Z_2) \nonu \\
&& + 
\frac{\bar{\theta}_{12}}{z_{12}}
 \left[ \frac{3(-6+c)(-1+c)}{(3+c)(-12+5c)} \alpha \overline{D} W +
\frac{3c(9+3c+2c^2)}{4(-1+c)(6+c)(-3+2c)} 
  \pa^2 \overline{D} T + \overline{D} V \right](Z_2)
\nonu \\
&& + 
\frac{\theta_{12}}{z_{12}}
 \left[ \frac{3(-6+c)(-1+c)}{(3+c)(-12+5c)} \alpha D W -
\frac{3c(9-3c+c^2)}{2(-1+c)(6+c)(-3+2c)} 
  \pa^2 D T - D V \right](Z_2)
\nonu \\
&& + 
\frac{\theta_{12}\bar{\theta}_{12}}{z_{12}}
 \left[- \frac{(-15+c)c}{(3+c)(-12+5c)} \alpha [D, \overline{D}] W +
\frac{(-18-3c-2c^2+2c^3)}{2(-1+c)(6+c)(-3+2c)} 
  \pa^3 T + 2 \pa V \right](Z_2)
\nonu \\
&& + (\mbox{Non-linear singular terms}) +\cdots,
\label{openolimit}
\eea
where the central charge is given by (\ref{centralcharge}) and the
self-coupling constant is given by (\ref{selfcoupling})
\bea
 c & = &  c(N,k) = \frac{3 N k}{ N +k +1}, \label{twoconsts} \\
 \alpha^2  & = & \alpha(N, k)^2 = 
\frac{3 (1+k)^2 (k-N)^2 (1+N)^2}{(-1+k) (-1+N) (1+2 k+N) (1+k+2 N)
  (-1-N+k (-1+3 N))}.
\nonu
\eea
One should see the linear structure in (\ref{openolimit}) for the
operator product expansion of  
the current with spins $(2, \frac{5}{2}, \frac{5}{2}, 3)$ and itself
in ${\cal  N}=2$ ${\cal W}_{N+1}$ algebra.
For the nonlinear terms, one has $TT(Z_2)$ term in the
$\frac{1}{z_{12}^2}$ term and the descendant fields arise in the
appropriate singular terms. One also has the nonlinear terms 
$T W(Z_2), T [D, \overline{D}]T(Z_2), T T
T(Z_2), \overline{D} T D T(Z_2)$ whose component fields appear in (\ref{spin3comp}). 

In principle, with the two values (\ref{twoconsts}), one can find
other higher spin currents. For given the higher spin current $W(Z)$ of spins $(2,
\frac{5}{2}, \frac{5}{2}, 3)$ in the ${\cal N}=2$ ${\cal W}_{N+1}$ algebra,  
one can construct the operator product expansion of this current and
itself.
By looking at the singular terms, one can read off the next higher
spin current, for example, $V(Z)$ of spins $(3, \frac{7}{2},
\frac{7}{2}, 4)$ given by (\ref{Vexpress}). 
Then one can continue to obtain the operator product
expansion between $W(Z_1)$ and $V(Z_2)$ in order to find other higher
spin current and so on \footnote{
Let us remind that for the bosonic case, the spin $3$-spin $3$
operator product expansion determines the 
spin $4$ current in the right hand side \cite{Ahn2011} up
to the overall normalization constant that can be fixed by the highest
singular term of spin $4$-spin $4$ operator product expansion.
Then one can compute the spin $3$-spin $4$ operator product expansion
and determine other higher spin current. For example, the spin $5$ current.  
The ${\cal N}=2$ ${\cal W}_5$ algebra should related to 
this bosonic $W_5$ algebra.
It would be interesting to find the structure constant for this
particular coeffcient in front of spin $5$ current in the right hand
side
and see whether this will coincide with the previous result by using
different method.}. 

According to the observation of \cite{GG1}, 
the original proposal in \cite{GG} should hold at finite $(N,k)$ and
one expects that the quantum deformation algebra of ${\cal N}=2$ 
${\cal W}_{\infty}^{\rm{cl}}[\lambda]$ in \cite{HP} should satisfy the
algebraic structure in (\ref{openolimit}), at finite $(N,k)$.

\section{The large $(N,k)$ 't Hooft limit of ${\cal N}=2$ ${\cal
    W}_{N+1}$ algebra}

We would like to describe the large $(N, k)$ limit for the operator
product expansion between the lowest higher spin current $W(Z_1)$ with spins
$(2, \frac{5}{2}, \frac{5}{2}, 3)$
and itself $W(Z_2)$. The large $(N,k)$ limit for fixed 't Hooft 
coupling constant $\la$ is given by
\bea
c(N,k) = \frac{3N k}{N+k+1} \longrightarrow 3(1-\la) N, \qquad \la \equiv \frac{N}{N+k}.
\label{limit}
\eea
Similarly, one also has 
the following limit for the self-coupling constant (\ref{selfcoupling}) 
\bea
\alpha(N,k)^2 \longrightarrow 
-\frac{(-1+2 \lambda )^2}{(-2+\lambda ) (1+\lambda )}.
\label{alphalimit1}
\eea

From the observations for $N=2$ and $N=4$ cases in previous
subsections, one expects that the operator product expansion, in the
large $(N,k)$ limit, together with (\ref{limit}) and
(\ref{alphalimit1}), 
takes the form 
\bea
&& W(Z_1) W(Z_2)  = 
\nonu \\
&&
\frac{1}{z_{12}^4} \,\, \frac{c(N,k)}{2} +\frac{\theta_{12}
  \bar{\theta}_{12}}{z_{12}^4}
\,\, 3 T(Z_2) +\frac{\bar{\theta}_{12}}{z_{12}^3} \,\, 3 \overline{D} T(Z_2)
-\frac{\theta_{12}}{z_{12}^3} \,\, 3 D T(Z_2) +\frac{\theta_{12} 
\bar{\theta}_{12}}{z_{12}^3} \,\, 3 \pa T(Z_2)
\nonu \\
&& + \frac{1}{z_{12}^2} \left[ 2 \alpha(N,k) W + [D,
  \overline{D}] T \right](Z_2) +\frac{\bar{\theta}_{12}}{z_{12}^2}
 \left[ \alpha(N,k) \overline{D} W +2
  \pa \overline{D} T \right](Z_2)
\nonu \\
&& +  \frac{\theta_{12}}{z_{12}^2}
 \left[ \alpha(N,k) D W -2
  \pa D T \right](Z_2) 
 + 
\frac{\theta_{12} \bar{\theta}_{12}}{z_{12}^2} \left[ 
\frac{3}{10} \alpha(N,k) [D, \overline{D} ] W 
 +  
\frac{3}{2} \pa^2 T + 3 V \right](Z_2)
\nonu \\
&& +  \frac{1}{z_{12}} \left[  
\alpha(N,k) \pa  W -\frac{1}{2} \pa [D,
  \overline{D}] T \right](Z_2) 
 + 
\frac{\bar{\theta}_{12}}{z_{12}}
 \left[ \frac{3}{5} \alpha(N,k) \overline{D} W +
\frac{3}{4} 
  \pa^2 \overline{D} T + \overline{D} V \right](Z_2)
\nonu \\
&& + 
\frac{\theta_{12}}{z_{12}}
 \left[ \frac{3}{5} \alpha(N,k) D W -
\frac{3}{4} 
  \pa^2 D T - D V \right](Z_2)
\nonu \\
&& + 
\frac{\theta_{12} \bar{\theta}_{12}}{z_{12}}
 \left[ -\frac{1}{5} \alpha(N,k) [D, \overline{D}] W +
\frac{1}{2} 
  \pa^3  T + 2 \pa V \right](Z_2)
\nonu \\
&& + \frac{1}{N} \mbox{(quadratic singular terms)} 
+ \frac{1}{N^2} \mbox{(cubic singular terms)} 
\nonu \\
&& + \frac{1}{N^3} \mbox{(quartic singular terms)} + \cdots,
\label{final}
\eea
where $c(N,k)$ and $\alpha(N,k)$ are the values after taking the
large $(N,k)$ limit, given in (\ref{limit}) and
(\ref{alphalimit1}) 
respectively.
At the linear order in the right hand side, 
we replace the fixed coupling constants (\ref{const1})
and 
(\ref{const2}) with the general coupling constant (\ref{selfcoupling})
and allow  to include the new ${\cal N}=2$ 
primary field $V(Z_2)$(and its descendant fields) (\ref{extra}) in the right
hand side of the operator product expansion (\ref{final}).
We list the component results of (\ref{final}) in the Appendix $G$.
Note that the $\pa [D, \overline{D}] T$ term in $\frac{\theta_{12} 
\bar{\theta}_{12}}{z_{12}^2}$ in (\ref{open2linear}) vanishes in this
limit and does not appear in (\ref{final}) also. 

One expects that the extra new composite fields 
$T^4(Z_2), T^2 W(Z_2), W^2(Z_2)$, and $TV(Z_2)$ with spins 
$(4, \frac{9}{2}, \frac{9}{2},5)$ should appear in the 
lowest singular term 
$\frac{\theta_{12} \bar{\theta}_{12}}{z_{12}}$ in (\ref{final}).
We have seen this feature in the ${\cal N}=2$ ${\cal W}_4$ algebra in
\cite{BW} although the full structure of the algebra is not given.
Also one sees the appearance of these new fields in the $AdS_3$ side. 
In \cite{HP}, the nonlinear terms in $(3.46)$ to $(3.53)$ contain 
these fields. For example, the $\frac{1}{k_{CS}^3}$ 
term corresponds to $T^4(Z_2)$ term and some of the
$\frac{1}{k_{CS}^2}$ terms contain $T^2 W(Z_2)$ term and so on.
Note that the Chern-Simon level $k_{CS}$ behaves as $N$ in the large
$N$ 't Hooft limit.

\section{ Comparison with 
the ${\cal N}=2$ classical ${\cal W}_{\infty}^{\rm{cl}}[\la]$ algebra
of the bulk theory }

One identifies the currents in the Kazama-Suzuki model with the higher
spin fields in ${\cal W}_{\infty}^{\rm{cl}}[\la]$ introduced in
\cite{HP} as follows:
\bea
T(z) & \longleftrightarrow & a_{\frac{3}{2}} \sim W_{1,HP}^{-}(z), \nonu \\ 
( D T +\overline{D} T) (z) & \longleftrightarrow & 
\psi_{\frac{3}{2}} \sim G_{2,HP}^{-}(z),
\nonu \\
( D T -\overline{D} T) (z) & \longleftrightarrow & \psi_{2}  
\sim G_{2,HP}^{+}(z), \nonu \\
-\frac{1}{2} [D, \overline{D}] T(z) & \longleftrightarrow & a_2
\sim W_{2,HP}^{+}(z),
\nonu \\
W(z) & \longleftrightarrow & a_{\frac{5}{2}} \sim W_{2,HP}^{-}(z), \nonu \\ 
( D W +\overline{D} W) (z) & \longleftrightarrow &
\psi_{\frac{5}{2}} \sim G_{3,HP}^{-}(z),
\nonu \\
( D W -\overline{D} W) (z) & \longleftrightarrow & \psi_{3}  \sim G_{3,HP}^{+}(z), \nonu \\
-\frac{1}{2} [D, \overline{D}] W(z) & \longleftrightarrow & a_3
\sim W_{3,HP}^{+}(z),
\nonu \\
V(z) & \longleftrightarrow & a_{\frac{7}{2}} \sim W_{3,HP}^{-}(z), \nonu \\ 
( D V +\overline{D} V) (z) & \longleftrightarrow &
\psi_{\frac{7}{2}} \sim G_{4,HP}^{-}(z),
\nonu \\
( D V -\overline{D} V) (z) & \longleftrightarrow & \psi_{4}  \sim G_{4,HP}^{+}(z), \nonu \\
-\frac{1}{2} [D, \overline{D}] V(z) & \longleftrightarrow & a_4
\sim W_{4,HP}^{+}(z).
\label{HPrelations}
\eea
We also present the CFT fields with $HP$ index in the last entry in
order to specify them from \cite{HP}. 

In order to obtain the $AdS_3$ result, the normalization factor should
occur in the operator product expansion of spin $2$ current and itself 
\bea
\beta(N,k)^2  & = & \frac{(-1+k) (-1+N) (1+2 k+N) (1+k+2 N)}{3 (1+k+N)^2
  (-1-k-N+3 k N)} \nonu \\
& \longrightarrow & -\frac{2}{9} (-1+\lambda_{HP} ) (1+2 \lambda_{HP} )=
-\frac{1}{9} (-2+\lambda ) (1+\lambda ), \,\,\,\, 2\la_{HP} = \la,
\label{betalimit}
\eea
where we also present the large $(N,k)$ limit (\ref{limit}).
This expression is a genealization of \cite{Ito} where the
$\beta(N=2,k)$
was found for fixed $N=2$ case. 
Similarly, one also has 
the following limit for the self-coupling constant
(\ref{selfcoupling})
as before 
\bea
\alpha(N,k)^2 \longrightarrow -\frac{(-1+4 \lambda_{HP} )^2}{2 (-1+\lambda_{HP} ) (1+2
  \lambda_{HP} )}=
-\frac{(-1+2 \lambda )^2}{(-2+\lambda ) (1+\lambda )}.
\label{alphalimit}
\eea

From the operator product expansion in Appendix $G$
and the following relations between our currents and the field
contents in \cite{HP}
\bea
W(z) \equiv \frac{1}{\beta(N,k)} W_2^{-}(z), \qquad
-\frac{1}{2} [D, \overline{D}] T(z) \equiv W_{2,HP}^{+}(z),
\label{relations}
\eea
one rewrites the  equation (\ref{ope345}), together with (\ref{HPrelations}),
as 
\bea
 W_{2,HP}^{-}(z) W_{2,HP}^{-}(w)  & = &  \frac{1}{(z-w)^4} \,\, \frac{1}{2}
c(N,k)
\beta(N,k)^2
\nonu \\
&+ & \frac{1}{(z-w)^2} \beta(N,k)^2 \left[
  \frac{2 \alpha(N,k)}{\beta(N,k)} W_{2,HP}^{-} + 2 W_{2,HP}^{+} \right](w)  \nonu \\
&+ &   \frac{1}{(z-w)} \beta(N,k)^2 \left[ 
\frac{\alpha(N,k)}{\beta(N,k)} \pa W_{2,HP}^{-} + \pa W_{2,HP}^{+} \right](w) 
 \nonu \\
& + &   \frac{1}{N} \mbox{(Non-linear singular terms)} +  \cdots
\nonu \\
& \longrightarrow &  
\frac{1}{(z-w)^4} \,\, (-1)\frac{1}{3}(1-\la_{HP})(2\la_{HP}-1)(2\la_{HP}+1)N
\nonu \\
& + & \frac{1}{(z-w)^2} \left[
\frac{2}{3}(1-4\la_{HP})  W_{2,HP}^{-} - 
\frac{4}{9} (2\la_{HP}+1)(\la_{HP}-1) W_{2,HP}^{+} \right](w)
\nonu \\
& + &  \frac{1}{(z-w)} \left[
\frac{1}{3}(1-4\la_{HP})  \pa W_{2,HP}^{-} - 
\frac{2}{9} (2\la_{HP}+1)(\la_{HP}-1) \pa W_{2,HP}^{+} \right](w)
\nonu \\
& + &   \frac{1}{N} \mbox{(Non-linear singular terms)} +\cdots,
\label{spin2spin2limit}
\eea
where we use the large $(N,k)$ limits for $\alpha(N,k), \beta(N,k)$
and $c(N,k)$, (\ref{alphalimit}), (\ref{betalimit}) and (\ref{limit})
respectively
\footnote{A commutator relation for the modes $
  (W_{2,HP}^{\mp})_m$
 is as follows: $ [(W_{2,HP}^{-})_m,
(W_{2,HP}^{-})_n]=
\beta(N,k)^2 (m-n) \left[  \frac{\alpha(N,k)}{\beta(N,k)} 
(W_{2,HP}^{-})_{m+n}  + (W_{2,HP}^{+})_{m+n} \right] + \beta(N,k)^2 
\frac{c(N,k)}{12} m (m^2-1) \delta_{m+n,0}+ \mbox{Nonlinear
  terms}$, where $ W_{2,HP}^{\mp} = \sum_{m \in {\bf Z}} \frac{
  (W_{2,HP}^{\mp})_m}{z^{m+2}}$. 
One sees the similar structure in \cite{Romans}.}.
This is exactly the same as the equation $(4.8)$ of \cite{HP}.
Then it is straightforward to change the above to the commutator and
agree with the $AdS_3$ result where one can use the identities
\bea
\beta(N,k)^2 \longrightarrow -N^B_{\frac{5}{2}} = - \frac{1}{3} N_3^B,
\qquad
N^B_{\frac{5}{2}} = \frac{2}{9}(-1+\la_{HP})(1+2\la_{HP})=\frac{1}{3} N_3^B,
\label{limitrelations}
\eea
where $N^B_{\frac{5}{2}}$ and $N_3^B$ in \cite{HP} are some normalization functions 
that depend on 't Hooft coupling constant and they appear in the
commutator relations in the $AdS_3$ side.
One might ask whether there exists a possibility for the existence of
a new primary field of spin in the $\frac{1}{(z-w)}$ term
(\ref{spin2spin2limit}).
If there is a new primary field in that singular term, one can 
change the arguments $z$ and $w$ and use the series expansion around $w$.
Then it turns out there is a minus sign for this primary field. This
implies that there is no extra new primary field of spin $3$.

By using the identification 
\bea
-\frac{1}{2} [D, \overline{D}] W(z) \equiv \frac{1}{\beta(N,k)} W_{3,HP}^{+}(z), \qquad
V(z) \equiv \frac{1}{\beta(N,k)^2} W_{3,HP}^{-}(z),
\label{defV}
\eea
and (\ref{relations}),
one also computes the large $(N,k )$ limit for the operator product
expansion between the spin $2$ current and the spin $3$ current, from
(\ref{ope346}),  as follows:
\bea
W_{2,HP}^{-}(z)  W_{3,HP}^{+}(w) & = &  
\frac{1}{(z-w)^4} \,\, 3 \beta(N,k)^2 W_{1,HP}^{-}(w) \nonu \\
& + & \frac{1}{(z-w)^2} \,\, \beta(N,k)^2 
\left[\frac{3}{5} \frac{\alpha(N,k)}{\beta(N,k)}  W_{3,HP}^{+} + 
\frac{3}{\beta(N,k)^2} W_{3,HP}^{-} \right](w) 
\nonu \\
& + &  
\frac{1}{(z-w)} \,\, \beta(N,k)^2 \left[\frac{1}{5} \frac{\alpha(N,k)}{\beta(N,k)} \pa
  W_{3,HP}^{+} 
+ \frac{1}{\beta(N,k)^2} \pa W_{3,HP}^{-} \right](w) \nonu \\
&+ &   \frac{1}{N} \mbox{(Non-linear singular terms)} + \cdots
\nonu \\
& \longrightarrow &
\frac{1}{(z-w)^4}  (-1) \frac{2}{3} (2\la_{HP}+1)(\la_{HP}-1) W_{1,HP}^{-}(w) \nonu \\
&+ & 
\frac{1}{(z-w)^2} \left[
 \frac{1}{5}(1-4\la_{HP})  W_{3,HP}^{+} + 3 W_{3,HP}^{-} \right](w)
\nonu \\
& + &  \frac{1}{(z-w)} \left[
 \frac{1}{15}(1-4\la_{HP})  \pa W_{3,HP}^{+} + 
\pa W_{3,HP}^{-} \right](w)
\nonu \\
& + & \frac{1}{N} \mbox{(Non-linear singular terms)} + \cdots.
\label{ope23}
\eea
One easily sees that this (\ref{ope23}) agrees with 
the equation $(3.46)$ in \cite{HP} at the linear order
\footnote{This can be written in terms of modes as follows:
$[(W_{2,HP}^{-})_m,  (W_{3,HP}^{+})_n]=\frac{\beta(N,k)^2}{2} m(m+1)
(W_{1,HP}^{-})_{m+n}-
\frac{1}{5} \beta(N,k)^2 (2m-n) \frac{\alpha(N,k)}
{\beta(N,k)} (W_{3,HP}^{+})_{m+n} + (2m-n) (W_{3,HP}^{-})_{m+n}+
\mbox{Nonlinear terms}$, which can be compared to \cite{Romans}.}.
For example, the relative
coefficient $\frac{1}{3}$ on the descendant field $ \pa W_{3,HP}^{-}$
can be obtained from   \cite{Dotsenko,Ahn92}
\bea
\frac{h_p + h_m -h_n }{2 h_p} = \frac{3 + 2 -3}{2
  \times 3} =\frac{1}{3}.
\nonu
\eea
It is not strange that there is no descendant field for the
$W_{1,HP}^{-}(w)$
because according to the counting of (\ref{formula}), the numerator
becomes zero($h_m=2, h_n=3$ and $h_p=1$). This implies that the
coefficient for the descendant field $\pa W_{1,HP}^{-}(w)$ vanishes and
there is no such term in the $\frac{1}{(z-w)^3}$ term in (\ref{ope23}).

Let us present the final bosonic operator product expansion between
the spin $3$ current and itself, from (\ref{ope347}),  where we use (\ref{limitrelations})
\bea
&& W_{3,HP}^{+}(z)  W_{3,HP}^{+}(w) 
 =  \frac{1}{(z-w)^6} \,\, \frac{5}{2} c(N,k) \beta(N,k)^2 \nonu \\
&& +  \frac{1}{(z-w)^4} \,\,
 \beta(N,k)^2
\left[ 3 \frac{\alpha(N,k)}{\beta(N,k)} W_{2,HP}^{-}
  +15 W_{2,HP}^{+} \right](w) \nonu \\
&& +   \frac{1}{(z-w)^3} \,\, \beta(N,k)^2 \left[ \frac{3}{2} \frac{\alpha(N,k)
 }{\beta(N,k)} \pa W_{2,HP}^{-}
  +\frac{15}{2} \pa  W_{2,HP}^{+} \right](w) \nonu \\
&& + 
\frac{1}{(z-w)^2} \,\, \beta(N,k)^2 
\left[ \frac{9}{20} \frac{\alpha(N,k) }{\beta(N,k)} \pa^2 W_{2,HP}^{-}
+\frac{9}{4} \pa^2 W_{2,HP}^{+} + \frac{4}{\beta(N,k)^2}  W_{4,HP}^{+}  \right](w)
\nonu \\
&& +   \frac{1}{(z-w)} \,\, \beta(N,k)^2 \left[  
\frac{1}{10} \frac{\alpha(N,k)}{\beta(N,k)} \pa^3 W_{2,HP}^{-}
+\frac{1}{2} \pa^3  W_{2,HP}^{+} + \frac{2}{\beta(N,k)^2} \pa W_{4,HP}^{+}
\right](w) 
\nonu \\
&& +   \frac{1}{N} \mbox{(Non-linear singular terms)} +\cdots
\nonu \\
&& \longrightarrow 
\frac{1}{(z-w)^6} (-1)\frac{5}{3}(1-\la_{HP})(2\la_{HP}-1)(2\la_{HP}+1)N
\nonu \\
&& +  \frac{1}{(z-w)^4} \left[ (1-4\la_{HP})  W_{2,HP}^{-} 
- \frac{10}{3} (2\la_{HP}+1)(\la_{HP}-1) W_{2,HP}^{+}
\right]
\nonu \\
&& +  \frac{1}{(z-w)^3} \left[ \frac{1}{2} (1-4\la_{HP})  \pa W_{2,HP}^{-} 
- \frac{5}{3} (2\la_{HP}+1)(\la_{HP}-1) \pa W_{2,HP}^{+}
\right]
\nonu \\
&& +  \frac{1}{(z-w)^2} \left[ \frac{3}{20} (1-4\la_{HP})  \pa^2 W_{2,HP}^{-} 
-\frac{1}{2}  (2\la_{HP}+1)(\la_{HP}-1) \pa^2 W_{2,HP}^{+} +4  W_{4,HP}^{+}
\right]
\nonu \\
&& +  \frac{1}{(z-w)} \left[ \frac{1}{30} (1-4\la_{HP})  \pa^3 W_{2,HP}^{-} 
- \frac{1}{9} (2\la_{HP}+1)(\la_{HP}-1) \pa^3 W_{2,HP}^{+} +2 \pa  W_{4,HP}^{+}
\right] 
\nonu \\
&& + \frac{1}{N} \mbox{(Non-linear singular terms)} +\cdots.
\label{ope33}
\eea
It is obvious that this equation (\ref{ope33})
should correspond to the equation $(3.47)$ of \cite{HP} \footnote{
One can express this as follows:
$[(W_{3,HP}^{+})_m, (W_{3,HP}^{+})_n] = \beta(N,k)^2 
\frac{c}{48} m(m^2-1)(m^2-4) \delta_{m+n,0} + \beta(N,k)^2 (m-n)
\left[\frac{1}{15}(m+n+3)(m+n+2) -
\frac{1}{6}(m+2)(n+2) \right]  \left[ \frac{3}{2} \frac{\alpha(N,k)}
{\beta(N,k)} (W_{2,HP}^{-})_{m+n} -\frac{15}{4} 
(W_{2,HP}^{+})_{m+n} \right] + 2  (m-n)
(W_{4,HP}^{-})_{m+n}+\mbox{Nonlinear terms}$.
Similarly, one can compare this with the corresponding equation in \cite{Romans}.}.
Also note that the relative coefficient function $\frac{1}{2}$ on $\pa
W_{4,HP}^{+}$ 
can be obtained
from the formula (\ref{formula}) by substituting $h_m = 3 = h_n$ and
$h_p=4$. The relative coefficients $1, \frac{1}{2}, \frac{3}{20}$, and
$\frac{1}{30}$, 
for the spin $2$ current in the
right hand side, are standard values in the well-known $W_3$
algebra. See, for example, the review paper \cite{BS}. 
The coefficient $\frac{3}{20}$ is nothing but
$\frac{1}{4}\frac{h_p+1}{2h_p+1}$
and this becomes $\frac{3}{20}$ at $h_p=2$ \cite{BPZ}.   

We also present the remaining $6$ operator product expansions, in the large
$(N,k)$ limit in (\ref{remain1}), (\ref{remain2}), (\ref{remain3}), 
(\ref{remain4}), (\ref{remain5}), and (\ref{remain6}). 
Due to the ${\cal N}=2$ supersymmetry(the current multiplets $T(Z)$
and $W(Z)$ and their operator product expansions can be
organized in manifest ${\cal N}=2$ superspace), compared to the
bosonic case, one could obtain much informations on the various operator
product expansions. In other words, for given operator product
expansion of ${\cal N}=2$ currents(only after this is determined by
other method, for example, Jacobi identity), there exist $16$ component
operator product expansions. Without any input for the ${\cal N}=2$
supersymmetry,
one should analyze all these operator product expansions 
separately \cite{Romans}.   
For example, for the ${\cal N}=2$ ${\cal W}_3$ algebra, by exploiting
the package of \cite{KT} with Jacobi identity, 
one can easily obtain the operator product
expansion for the higher spin current in ${\cal N}=2$ superspace.

\section{Conclusions and outlook }

We have constructed the ${\cal N}=2$ current with spins $(2,
\frac{5}{2}, \frac{5}{2}, 3)$ in (\ref{superspin2}) and the
self-coupling constant in (\ref{const2}) in ${\cal N}=2$ ${\cal W}_5$ algebra. 
We also have found the extra singular terms in the 
operator product expansion in
(\ref{extra}) which were not present in ${\cal N}=2$ ${\cal W}_3$ algebra.   
By observing the self-coupling constant (\ref{selfcoupling}) 
which depends on $(N,k)$ explicitly, 
the large $(N,k)$ limit of ${\cal N}=2$ ${\cal W}_{N+1}$ algebra
contains the particular operator product expansion given in (\ref{final}).
We have identified this with the corresponding ${\cal N}=2$ classical
${\cal W}_{\infty}^{\rm{cl}}[\lambda]$ algebra in the bulk.

$\bullet$ It is an immediate question to ask how one obtains the 
higher spin current including (\ref{n2terms}) and (\ref{superspin2}) for ${\cal N}=2$ 
${\cal W}_{N+1}$ algebra. From the structure of (\ref{superspin2}),
one can try to write down the correct ansatz for the possible
terms(one might add a few extra terms which are not present for $N=2$
or $N=4$ case)
and then apply to the
two conditions 1) and 2) in (\ref{KWvanishing}) and (\ref{TW}). It is
nontrivial to find the identities for the multiple products between 
the structure constants in the complex basis and to 
collect the independent fields in each singular term.   
These are necessary to check the right singular structures.

$\bullet$ It would be interesting to obtain 
the quantum ${\cal N}=2$ ${\cal W}_{\infty}^{\rm{qu}}[\lambda]$
algebra which
is a deformation of the classical ${\cal N}=2$ 
${\cal W}_{\infty}^{\rm{cl}}[\lambda]$ algebra.  
For the KS model side, once we complete all the operator product
expansions at least ${\cal N}=2$ ${\cal W}_5$ algebra, then this
algebra should provide all the informations on the quantum ${\cal N}=2$ 
${\cal W}_{\infty}^{\rm{qu}}[\lambda]$
algebra, along the line of \cite{GG1}.  
The ${\cal W}_{\infty}^{\rm{cl}}[\lambda]$ for the bosonic case 
was found in \cite{GH, GHJ} and the corresponding quantum algebra has
been  studied in \cite{GG1}. See also \cite{BBSS1,Ahn2011}.
We expect that there are extra linear terms for the nonlinear 
composite currents in (\ref{final}).
From the observation \cite{BW1}, as we take $c \rightarrow \infty$ in
the quantum operator product expansion, any composite field(product of
$n$ fields) where the $c$'s power in the denominator is greater than $(n-1)$   
will disappear in the classical limit.
For example, the standard spin $3$-spin $3$ operator product expansion 
has the nonlinear term $\Lambda(w) \equiv 
TT(w) -\frac{3}{10} \pa^2 T(w)$ with
coefficient function $\frac{32}{22+5c}$ in $\frac{1}{(z-w)^2}$ term as
well as $\frac{3}{10} \pa^2 T(w)$ term \cite{BS}.
In the $c \rightarrow \infty$ limit, the $\pa^2 T(w)$ term in the
$\Lambda(w)$ 
vanishes 
while the $T T(w)$ term survives.
In quantum theory, the extra term like as $\pa^2 T(w)$ in the
$\Lambda(w)$
exists.
On the other hand, it is an open problem to obtain the bosonic
subalgebra(how to one gets the bosonic 
$W_5$ algebra) or ${\cal N}=1$ subalgebra for the algebra we have described,
along the line of \cite{Romans}.

$\bullet$ According the classification for the KS model \cite{KSNPB},
there exists the following coset model also
\bea
\frac{SO(N+2)}{SO(N) \times SO(2)}, \qquad c(N,k) =\frac{3Nk}{N+k}.
\nonu
\eea
It would be interesting to find the higher spin currents for this
model and see how they
arise as an ${\cal N}=2$ nonlinear algebra.
Once we construct the complex basis for the group $SO(N+2)$, then the
current algebra similar to (\ref{basicOPE}) should exist. Only the
structure constants and dual Coxeter number can change.
Then the standard Sugawara construction can follow similarly, along
the line of \cite{Ahn2012}. See also the relevant works in \cite{Ahn1106,GV}.

$\bullet$ As pointed out in \cite{HGPR}, it would be interesting to
construct the more supersymmetric higher spin $AdS_3$ supergravity 
dual to the  ${\cal N}=4$ superconformal coset model that can be
realized by the ${\cal N}=4$ current algebra for the supersymmetric
WZW model. 
As a first step, one can use the previous work of \cite{RASS} where
the ${\cal N}=4$ superconformal algebra(the spins for 
all the currents are less than or equal to $2$: the spin $2$ current,
four spin $\frac{3}{2}$ currents, seven spin $1$ currents and four
spin $\frac{1}{2}$ currents) 
can be written in terms of the
${\cal N}=2$ affine Kac-Moody currents. 
It is an open problem to construct the higher spin currents with spin
greater than $2$. In ${\cal N}=2$ superspace, one should have $T(Z)$
with spins $(1, \frac{3}{2}, \frac{3}{2}, 2)$ and $W(Z)$ with spins
$(2, \frac{5}{2}, \frac{5}{2}, 3)$ as well as the extra  primary currents. 
For the minimal extension of ${\cal N}=2$ ${\cal W}_3$ algebra, 
the number of these extra currents is equal to $2$. One of them corresponds
to the ${\cal N}=4$ partner of $T(Z)$ and the other corresponds to the
${\cal N}=4$ partner of $W(Z)$.   
The spins for the ${\cal N}=2$ multiplets 
can be either $(\frac{3}{2}, 2, 2, \frac{5}{2})$ or
$({\frac{5}{2}, 3, 3, \frac{7}{2}})$.
The former is more preferable because the extension of $W_3$
current(the last component of $W(Z)$)
has its partner of spin $\frac{5}{2}$ in the context of \cite{ASS}. 
Note that the full ${\cal N}=4$ superconformal algebra is generated by 
the stress energy tensor $T(Z)$ with spins $(1, \frac{3}{2},
\frac{3}{2}, 2)$, two ${\cal N}=2$ currents with spins $(\frac{1}{2}, 1, 1,
\frac{3}{2})$
and a ${\cal N}=2$ 
current with spins $(0, \frac{1}{2}, \frac{1}{2}, 1)$.

$\bullet$ From the result of \cite{GH}, one expects that 
the linear structure in (\ref{final}) should have the higher spin
algebra in \cite{BVW1,BVW2}, although the explicit relations are not
given in this paper.
One cannot use their expressions directly because the currents or
generators are not primary fields. So in order to compare with our
results here, one should obtain the correct primary fields with
respect to the stress energy tensor.
Of course, the higher spin algebra is not a subalgebra of 
the ultimate quantum algebra but is a subalgebra in the $c \rightarrow
\infty$ limit.
In general, the ultimate quantum algebra does not contain higher spin
algebra as a subalgebra.

$\bullet$ It is an open problem to reconsider the previous analysis in 
\cite{CHR},
under the large $(N,k)$ limit, along the line of \cite{GG1}. 
This can be done only after the ${\cal N}=2$ quantum 
${\cal W}_{\infty}^{\rm{qu}}[\lambda]$ algebra is found.

\vspace{.7cm}

\centerline{\bf Acknowledgments}

We would like to thank the following people for 
correspondence on the following topics: R. Gopakumar on the current
status of the triality \cite{GG1}, Y. Hikida on the
supersymmetric version of higher spin algebra \cite{CHR}, 
S. Krivonos on his mathematica package
for ${\cal N}=2$ operator product expansions \cite{KT}, S. Odake on his paper \cite{Odake},
C. Peng on the asymptotic symmetry \cite{HP}
and M. Vasiliev on the super $W_{\infty}(\lambda)$ algebra \cite{BVW1,BVW2}. 
This work was supported by the Mid-career Researcher Program through
the National Research Foundation of Korea (NRF) grant 
funded by the Korean government (MEST) (No. 2009-0084601).

\newpage

\appendix

\renewcommand{\thesection}{\large \bf \mbox{Appendix~}\Alph{section}}
\renewcommand{\theequation}{\Alph{section}\mbox{.}\arabic{equation}}

\section{The ${\cal N}=2$ current algebra }

The on-shell current algebra in ${\cal N}=2$ superspace 
for the supersymmetric WZW model, with level $k$,  
on a group $G=SU(N+1)$ of even-dimension, can be written as \cite{HS,Ahn94,Ahn92}
\bea
 Q^A (Z_{1}) Q^B (Z_{2})   & = & -\tzb f_{\bar{A}\bar{B}}^{\;\;\;\;\bar{C}} Q^C(Z_2)
-\tzbb \frac{1}{(k+N+1)} f_{\bar{A} C}^{\;\;\;\;\bar{D}}
 f_{\bar{B} \bar{C}}^
{\;\;\;\;\bar{E}}
 Q^D Q^E(Z_2),
\nonu \\
 Q^{\bar{A}} (Z_{1}) Q^{\bar{B}} (Z_{2})   
& = & -\tz f_{A B}^{\;\;\;\;C} Q^{\bar{C}}(Z_2)
+\tzbb \frac{1}{(k+N+1)} f_{A \bar{C}}^{\;\;\;\;D}
 f_{B C}^
{\;\;\;\;E}
 Q^{\bar{D}} Q^{\bar{E}}(Z_2),
\nonu \\
 Q^A (Z_{1}) Q^{\bar{B}} (Z_{2})  & = &
 \tzzbb \frac{1}{2} \left[(k+N+1) \delta^
{A \bar{B}}  + 
f_{\bar{A} C}^{\;\;\;\;\bar{D}} f_{B \bar{C}}^{\;\;\;\;D} 
\right]
  -\frac{1}{z_{12}} (k+N+1) \delta^{A \bar{B}} \nonu \\
 & & - \tz f_{\bar{A} B}^{\;\;\;\;\bar{C}} Q^{C}(Z_2)-
   \tzb f_{\bar{A} B}^{\;\;\;\;C}  Q^{\bar{C}}(Z_2) 
  \nonu \\
&&  - \tzbb \left[ f_{\bar{A} B}^{\;\;\;\;\bar{C}} \overline{D} Q^C 
+ \frac{1}{(k+N+1)}  f_{\bar{A} C}^{\;\;\;\;\bar{D}} f_{B \bar{C}}^{\;\;\;\;E}
 Q^D Q^{\bar{E}} 
 \right](Z_2),
\label{OPEQQ}
\eea
where the nonlinear constraints are given by \cite{HS,Ahn94}
\bea
D Q^A = -\frac{1}{2(k+N+1)} f_{\bar{A} B}^{\;\;\;\;\bar{C}} Q^B Q^C,
\qquad
\overline{D} Q^{\bar{A}} = 
-\frac{1}{2(k+N+1)} f_{A\bar{B}}^{\;\;\;\;C} Q^{\bar{B}} Q^{\bar{C}}.
\label{constraint}
\eea
The Jacobi identities of the algebra (\ref{OPEQQ}) 
are satisfied under the constraints (\ref{constraint}).
The operator product expansion 
$Q^{\bar{A}} (Z_{1}) Q^{B} (Z_{2})$ can be obtained 
from the third equation of (\ref{OPEQQ}). 
The stress energy tensor can be written as 
\bea
T(Z) = -\frac{1}{(k+N+1)} \delta_{A\bar{B}} Q^A Q^{\bar{B}}(Z) +
\frac{1}{(k+N+1)}\left[ \delta_{B\bar{C}} f_{\bar{A} \bar{B}}^{\;\;\;\;\bar{C}} 
D Q^{\bar{A}} +  \delta_{B\bar{C}} f_{A \bar{B}}^{\;\;\;\;\bar{C}} 
\overline{D} Q^{A} \right].
\label{Sugawara}
\eea
In the section $2$, we rewrite the equations (\ref{OPEQQ}),
(\ref{constraint}) 
and (\ref{Sugawara}) in manifest way of 
the subgroup $H$ and the coset $\frac{G}{H}$. 

\section{The ${\cal N}=2$ current algebra (\ref{basicOPE}) in the
  component approach }

The $22$ operator product expansions, by expanding the 
above operator product expansions (\ref{OPEQQ}) or (\ref{basicOPE}) into the component, 
can be summarized by
\bea
K^m(z) \overline{D} K^n(w) & = & -\frac{1}{(z-w)} f_{\bar{m} 
\bar{n}}^{\;\;\;\;\bar{p}} K^p(w) + \cdots,
\nonu \\
\overline{D} K^m (z) \overline{D} K^n(w) & = & -\frac{1}{(z-w)} f_{\bar{m}
  \bar{n}}^
{\;\;\;\;\bar{p}} \overline{D} K^p(w) + \cdots,
\nonu \\
K^m(z) \overline{D} J^a(w) & = & -\frac{1}{(z-w)} f_{\bar{m}
  \bar{a}}^{\;\;\;\;\bar{b}}
J^b(w) + \cdots,
\nonu \\
\overline{D} K^m (z) J^a(w) & = & -\frac{1}{(z-w)} f_{\bar{m} \bar{a}}^
{\;\;\;\;\bar{b}} J^b(w) +\cdots,
\nonu \\
\overline{D} K^m(z) \overline{D} J^a(w) & = & -\frac{1}{(z-w)} f_{\bar{m}
  \bar{a}}^{\;\;\;\; \bar{b}} \overline{D} J^b(w) +\cdots,
\nonu \\
K^{\bar{m}} (z) D K^{\bar{n}} (w) & = & -\frac{1}{(z-w)} f_{m 
n}^{\;\;\;\; p} K^{\bar{p}}(w) + \cdots,
\nonu \\
D K^{\bar{m}} (z) D K^{\bar{n}} (w) & = & -
\frac{1}{(z-w)} f_{m
 n}^
{\;\;\;\; p} D K^{\bar{p}} (w) + \cdots,
\nonu \\
K^{\bar{m}} (z) D J^{\bar{a}} (w) & = & -\frac{1}{(z-w)} 
f_{m a}^{\;\;\;\; b}
J^{\bar{b}}(w) + \cdots,
\nonu \\
D K^{\bar{m}} (z) J^{\bar{a}} (w) & = & 
-\frac{1}{(z-w)} f_{m a}^
{\;\;\;\; b} J^{\bar{b}} (w) +\cdots,
\nonu \\
D K^{\bar{m}} (z) D J^{\bar{a}} (w) & = & 
-\frac{1}{(z-w)} f_{m
  a}^{\;\;\;\; b} D J^{\bar{b}} (w) +\cdots,
\nonu \\
K^m(z) K^{\bar{n}}(w) & = & -\frac{1}{(z-w)} (k+N) \delta^{m\bar{n}}+
\cdots,
\nonu \\
K^{m} (z) D K^{\bar{n}} (w) & = & -\frac{1}{(z-w)} f_{\bar{m} 
n}^{\;\;\;\; \bar{p}} K^{p}(w) + \cdots,
\nonu \\
\overline{D} K^{m} (z) K^{\bar{n}} (w) & = & -
\frac{1}{(z-w)} f_{\bar{m}
 n}^
{\;\;\;\; p}  K^{\bar{p}} (w) + \cdots,
\nonu \\
\overline{D} K^{m} (z) D K^{\bar{n}} (w) & = & \frac{1}{(z-w)^2}
\frac{1}{2} \left[2(k+N+1) \delta^{m\bar{n}} + f_{\bar{m} p}^{\;\;\;\;
\bar{q}} f_{n\bar{p}}^{\;\;\;\;q} \right] \nonu \\
&  - &
\frac{1}{(z-w)} \left[ f_{\bar{m}
 n}^
{\;\;\;\; p} D K^{\bar{p}} +f_{\bar{m}
 n}^
{\;\;\;\; \bar{p}} \overline{D} K^{p}
 + \frac{1}{k+N+1} f_{\bar{m} p}^{\;\;\;\;\bar{q}} f_{n\bar{p}
}^{\;\;\;\;r} K^q K^{\bar{r}} \right] (w)  + \cdots,
\nonu \\
\overline{D} J^{a} (z) D J^{\bar{b}} (w) & = & \frac{1}{(z-w)^2}
\frac{1}{2} \left[2(k+N+1) \delta^{a\bar{b}} + 
f_{\bar{a} m}^{\;\;\;\;
\bar{c}} f_{b\bar{m}}^{\;\;\;\;c} + f_{\bar{a} c}^{\;\;\;\;\bar{m}} 
f_{b\bar{c}}^{\;\;\;\; m}\right] \nonu \\
&  - &
\frac{1}{(z-w)} \left[ f_{\bar{a}
 b}^
{\;\;\;\; \bar{m}} \overline{D} K^{m} +f_{\bar{a}
 b}^
{\;\;\;\; m} D K^{\bar{m}}
 \right. \nonu \\
& + & \left.  \frac{1}{k+N+1} \left( 
f_{\bar{a} m}^{\;\;\;\;\bar{c}} f_{b\bar{m}
}^{\;\;\;\;d} J^c J^{\bar{d}}  + f_{\bar{a}c}^{\;\;\;\;\bar{m}}
f_{b\bar{c}}^{\;\;\;\;n} K^m K^{\bar{n}} \right) \right] (w)  + \cdots,
\nonu \\
J^a(z) J^{\bar{b}}(w) 
& = & -\frac{1}{(z-w)} (k+N+1) \delta^{a\bar{b}}+
\cdots,
\nonu \\
J^{a} (z) D J^{\bar{b}} (w) & = & -\frac{1}{(z-w)} 
f_{\bar{a} b}^{\;\;\;\; \bar{m}}
K^{m}(w) + \cdots,
\nonu \\
\overline{D} J^{a} (z) J^{\bar{b}} (w) & = & 
-\frac{1}{(z-w)} f_{\bar{a} b}^
{\;\;\;\; m} K^{\bar{m}} (w) +\cdots,
\nonu \\
\overline{D} K^{m} (z) J^{\bar{a}} (w) & = & 
-\frac{1}{(z-w)} f_{\bar{m} a}^
{\;\;\;\; b} J^{\bar{b}} (w) +\cdots,
\nonu \\
\overline{D} K^{m} (z) D J^{\bar{a}} (w) & = & \frac{1}{(z-w)^2}
\frac{1}{2} f_{\bar{m} n}^{\;\;\;\;
\bar{p}} f_{a \bar{n}}^{\;\;\;\;p}  \nonu \\
&  - &
\frac{1}{(z-w)} \left[ f_{\bar{m}
 a}^
{\;\;\;\; b} D J^{\bar{b}} 
 + \frac{1}{k+N+1} f_{\bar{m} p}^{\;\;\;\;\bar{q}} f_{a \bar{p}
}^{\;\;\;\;b} K^q J^{\bar{b}} \right] (w)  + \cdots,
\nonu \\
J^{a} (z) D K^{\bar{m}} (w) & = & -\frac{1}{(z-w)} 
f_{\bar{a} m}^{\;\;\;\; \bar{b}}
J^{b}(w) + \cdots,
\nonu \\
\overline{D} J^{a} (z) D K^{\bar{m}} (w) & = &   - 
\frac{1}{(z-w)} \left[ f_{\bar{a}
 m}^
{\;\;\;\; \bar{b}} \overline{D} J^{b} 
 + \frac{1}{k+N+1} f_{\bar{a} p}^{\;\;\;\;\bar{b}} f_{m \bar{p}
}^{\;\;\;\;n} J^b K^{\bar{n}} \right] (w)  + \cdots.
\label{compexpression}
\eea
Here we use the component currents (\ref{components}) in order not to
introduce many different notations for various fields with different spins. 
Of course, the remaining $14(=36-22)$ operator product expansions 
do not have any singular terms
\bea
&& K^m (z) K^n(w)  =  0, \qquad K^m (z) J^a(w)=0, \qquad 
K^m (z) J^{\bar{a}}(w)=0, \qquad K^m (z) D J^{\bar{a}}(w)=0, 
\nonu \\
&& J^a (z) J^b(w)  =  0, \qquad J^a (z) \overline{D} J^b(w)=0, \qquad 
J^a (z) K^{\bar{m}}(w)=0, \qquad \overline{D} J^a (z) \overline{D} J^b
(w)=0, 
\nonu \\
&& \overline{D} J^a (z) K^{\bar{m}} (w)  =  0, \qquad K^{\bar{m}} (z)
K^{\bar{n}} (w)=0, \qquad 
K^{\bar{m}} (z) J^{\bar{a}}(w)=0, \qquad J^{\bar{a}} (z)  J^{\bar{b}}(w)=0, 
\nonu \\
&& J^{\bar{a}} (z) D J^{\bar{b}}(w)  =  0, \qquad D J^{\bar{a}} (z) D J^{\bar{b}}(w)=0.
\nonu 
\eea 
Due to the limitation of \cite{KT}, we compute the operator product
expansions for $N=4$ via the component approach given in \cite{Thielemans}.
Note that the $k$-dependent terms in $\overline{D} K^m (z) D
K^{\bar{n}}(w)$ and $\overline{D} J^a(z) D J^{\bar{b}}(w)$
(\ref{compexpression}) 
are the
same as $(k+N+1)$.
This was used in (\ref{centralcharge}).

\section{The operator product expansion (\ref{TW}) in the component  approach }

One writes the operator product expansion (\ref{TW}), from (\ref{compstress}) and (\ref{n2W}),
in terms of components as follows:
\bea
T(z) D W(w) & = & \frac{1}{(z-w)} D W(w) + \cdots,
\nonu \\
T(z) \overline{D} W(w) & = & -\frac{1}{(z-w)} \overline{D} W(w) + \cdots,
\nonu \\
T(z) [ D, \overline{D} ] W(w) & = & -\frac{1}{(z-w)^2} 4 W(w) + \cdots,
\nonu \\
D T(z)  W(w) & = & -\frac{1}{(z-w)} D W(w) + \cdots,
\nonu \\
D T(z) \overline{D} W(w) & = & \frac{1}{(z-w)^2} 2 W(w) +
\frac{1}{(z-w)} 
\frac{1}{2} \left[ -[ D, \overline{D} ] W +\pa W \right](w) + \cdots,
\nonu \\
D T(z) [D, \overline{D}] W(w) & = & - \frac{1}{(z-w)^2} 5 D W(w) -
\frac{1}{(z-w)} \pa D W(w) + \cdots,
\nonu \\
\overline{D} T(z)  W(w) & = & \frac{1}{(z-w)} \overline{D} W(w) + \cdots,
\nonu \\
\overline{D} T(z) D W(w) & = & - \frac{1}{(z-w)^2} 2 W(w) -
\frac{1}{(z-w)} 
\frac{1}{2} \left[ [ D, \overline{D} ] W +\pa W \right](w) + \cdots,
\nonu \\
\overline{D} T(z) [ D, \overline{D} ] W(w) & = & - \frac{1}{(z-w)^2} 5 
\overline{D} W(w) -
\frac{1}{(z-w)} \pa \overline{D} W(w) + \cdots,
\nonu \\
(-1) \frac{1}{2} \left[ D, \overline{D} \right] T(z) W(w) & = &  \frac{1}{(z-w)^2} 2 
 W(w) +
\frac{1}{(z-w)}  \pa  W(w) + \cdots,
\nonu \\
(-1) \frac{1}{2} \left[ D, \overline{D} \right] T(z) D W(w) & = &  
\frac{1}{(z-w)^2} \frac{5}{2} 
D W(w) +
\frac{1}{(z-w)}  \pa D W(w) + \cdots,
\nonu \\
(-1) \frac{1}{2} \left[ D, \overline{D} \right] T(z) \overline{D} W(w)
& = & 
 \frac{1}{(z-w)^2} \frac{5}{2} 
\overline{D} W(w) +
\frac{1}{(z-w)}  \pa \overline{D} W(w) + \cdots,
\nonu \\
(-1) \frac{1}{2} 
\left[ D, \overline{D} \right] T(z) [ D, \overline{D} ] W(w) & = &  \frac{1}{(z-w)^2} 3 
 [ D, \overline{D} ] W(w) +
\frac{1}{(z-w)}  \pa [ D, \overline{D} ] W(w) + \cdots.
\label{TWcomp}
\eea
Note that the remaining three operator product expansions 
$T(z) W(w), D T(z) D W(w)$, and $\overline{D} T(z) \overline{D} W(z)$ 
do not have any singular terms.
It is obvious, from the last four equations of (\ref{TWcomp}), 
that the component fields (\ref{n2W}) are primary with respect to
the stress energy tensor.

\section{The coefficient functions in the primary field in
  ${\cal N}=2$ ${\cal W}_3$ algebra }

In (\ref{n2terms}), we write down the spin $2$ current with various
contracted terms. The coefficient functions are given by
\bea
a_1 & = & -1,\qquad a_2 = -1,\nonu \\
a_3  & = & \frac{(15-7 k-2 k^2)}{(-6+10 k)}, \qquad
a_4 = -3-k,
\nonu \\
a_5 & = & -3-k, \qquad 
a_6= -\frac{2 i \sqrt{3} \left(-6+k+k^2\right)}{(-3+5
  k)},
\nonu \\
a_7 & = & 
\frac{2 i \sqrt{3} \left(-6+k+k^2\right)}{(-3+5 k)},
\qquad 
a_8 = \frac{k \left(15+8 k+k^2\right)}{(-3+5
  k)},\nonu \\
a_9 & = &  -3-k, \qquad
a_{10} = \frac{3 \left(-6-5 k+2
    k^2+k^3\right)}{(6-10 k)},
\nonu \\
a_{11} & = & \frac{3 \left(-6-5 k+2
    k^2+k^3\right)}{(-6+10 k)}, \qquad
a_{12} = \frac{1}{2}(3+k), \nonu \\
a_{13} & = &
-\frac{1}{2}(3+k), \qquad
a_{14} = \frac{k \left(15+8 k+k^2\right)}{(-6+10 k)},\nonu \\
a_{15} & = &
\frac{k \left(15+8 k+k^2\right)}{(-6+10 k)}, \qquad
a_{16} = -\frac{1}{2}(3+k),\nonu \\
a_{17} & = &
-\frac{1}{2}(3+k), \qquad
a_{18} = -\frac{k \left(15+8 k+k^2\right)}{(-3+5 k)},
\nonu \\
a_{19} & = &
3+k, \qquad 
a_{20} = \frac{(3+k) \left(-3+\left(5+5 i \sqrt{3}\right) k+i
    \sqrt{3} k^2\right)}{(-6+10 k)},
\nonu \\
a_{21} & = & \frac{(3+k) \left(3+5 i
    \left(i+\sqrt{3}\right) k+i \sqrt{3} k^2\right)}{(-6+10 k)},
\qquad 
a_{22} =
\frac{(27+6 k-k^2)}{(-6+10 k)},\nonu \\
a_{23} & = &  \frac{(27+6 k-k^2)}{(-6+10
  k)}, \qquad 
a_{24} = 
\frac{(-27-6 k+k^2)}{(-3+5 k)}.
\label{coeffn2one}
\eea

The other way to express the spin $2$ current with spins $(2,
\frac{5}{2}, \frac{5}{2}, 3)$ is to take the ansatz for $N=4$ choice
given in (\ref{superspin2}).
By requiring the two conditions (\ref{KWvanishing}) and (\ref{TW}), one
gets the following coefficient functions which depend on the level $k$
explicitly as follows:
\bea
b_1 & = & -\frac{(-9+7 k+4 k^2)}{(3+k) (-3+5 k)},
\qquad
b_2= 1,
\nonu \\
b_3 & = &
-\frac{(5+k) (-3+2 k)}{2 (-3+5 k)}, \qquad
b_4 = \frac{k (5+k)}{2 (-3+5
  k)},
\nonu \\
b_5 & = & -\frac{k (5+k)}{2 (-3+5 k)}, \qquad
b_6 = -\frac{(-18+19 k+9
  k^2)}{2 (-3+5 k)}, \nonu \\
b_7 &=& -\frac{(-18+19 k+9 k^2)}{2 (-3+5 k)}, \qquad 
b_8 =
-\frac{k (5+k)}{2 (-3+5 k)}, \nonu \\
b_9 & = & \frac{k (3+k) (5+k)}{(-3+5
  k)}, \qquad 
b_{10} = -3-k,
\nonu \\
b_{11} & = & -\frac{-36-15 k+10 k^2+5 k^3}{4 (-3+5
  k)}, \qquad
b_{12} = \frac{(-36-15 k+10 k^2+5 k^3)}{4 (-3+5 k)},
\nonu \\
b_{13} & = &
\frac{3+k}{2}, \qquad 
b_{14} = \frac{(1+k) \left(18+3 k+k^2\right)}{4 (-3+5
  k)}, \nonu \\
b_{15} & = & \frac{k (3+k) (5+k)}{4 (-3+5 k)}, \qquad
b_{16} = \frac{k (3+k)
  (5+k)}{4 (-3+5 k)}, \nonu \\
b_{17} & = & \frac{1}{2} (-3-k), \qquad
b_{18} = \frac{(-3+k)
  (-2+k) (3+k)}{4 (-3+5 k)}, \nonu \\
b_{19} & = &  3+k, \qquad
b_{20} = -\frac{i
  \left(-i+\sqrt{3}\right) k (3+k) (5+k)}{2 (-3+5 k)}, \nonu \\
b_{21} & = & i
\sqrt{3}, \qquad
b_{22}  =  -\frac{2 (-2+k) (3+k)}{(-3+5 k)}, \nonu \\
b_{23} & = & 
-\frac{2 (-2+k) 
(3+k)}{-3+5 k}, \qquad
b_{24}  =  \frac{(-9+k) (3+k)}{6 (-3+5 k)}, \nonu \\
b_{25} & = &  \frac{(-9+k) (3+k)}{3 (-3+5 k)}, \qquad
b_{26}  =  \frac{(-9+k) (3+k)}{6 (-3+5 k)},\nonu \\
b_{27} & = & 1.
\label{coeffn2two}
\eea

Now we present the operator product expansion of spin $2$ current and
itself \cite{Ahn94}, at the linearized level, as follows:
\bea
W(Z_1) W(Z_2) & = &
\frac{1}{z_{12}^4} \frac{c}{2} +\frac{\theta_{12}
  \bar{\theta}_{12}}{z_{12}^4}
3 T(Z_2) +\frac{\bar{\theta}_{12}}{z_{12}^3} 3 \overline{D} T(Z_2)
-\frac{\theta_{12}}{z_{12}^3} 3 D T(Z_2) +\frac{\theta_{12} 
\bar{\theta}_{12}}{z_{12}^3} 3 \pa T(Z_2)
\nonu \\
&+& \frac{1}{z_{12}^2} \left[ 2 \alpha W +\frac{c}{(-1+c)} [D,
  \overline{D}] T \right](Z_2) +\frac{\bar{\theta}_{12}}{z_{12}^2}
 \left[ \alpha \overline{D} W +\frac{(-3+2c)}{(-1+c)} 
  \pa \overline{D} T \right](Z_2)
\nonu \\
& + & \frac{\theta_{12}}{z_{12}^2}
 \left[ \alpha D W -\frac{(-3+2c)}{(-1+c)} 
  \pa D T \right](Z_2) 
 + 
\frac{\theta_{12} \bar{\theta}_{12}}{z_{12}^2} \left[ 
\frac{3(-8+c)}{2(-12+5c)} \alpha [D, \overline{D} ] W \right. \nonu \\ 
& + & \left. 
\frac{9c(-12+5c)}{4(-1+c)(6+c)(-3+2c)} \pa [D, \overline{D}] T +
\frac{3(18-15c+2c^2+2c^3)}{2(-1+c)(6+c)(-3+2c)} \pa^2 T \right](Z_2)
\nonu \\
& + & \frac{1}{z_{12}} \left[  
\alpha \pa  W -\frac{c}{2(-1+c)} \pa [D,
  \overline{D}] T \right](Z_2) \nonu \\
& + &
\frac{\bar{\theta}_{12}}{z_{12}}
 \left[ \frac{3(-6+c)(-1+c)}{(3+c)(-12+5c)} \alpha \overline{D} W +
\frac{3c(9+3c+2c^2)}{4(-1+c)(6+c)(-3+2c)} 
  \pa^2 \overline{D} T \right](Z_2)
\nonu \\
& + &
\frac{\theta_{12}}{z_{12}}
 \left[ \frac{3(-6+c)(-1+c)}{(3+c)(-12+5c)} \alpha D W -
\frac{3c(9-3c+c^2)}{2(-1+c)(6+c)(-3+2c)} 
  \pa^2 D T \right](Z_2)
\nonu \\
& + &
\frac{\theta_{12}\bar{\theta}_{12}}{z_{12}}
 \left[- \frac{(-15+c)c}{(3+c)(-12+5c)} \alpha [D, \overline{D}] W +
\frac{(-18-3c-2c^2+2c^3)}{2(-1+c)(6+c)(-3+2c)} 
  \pa^3 T \right](Z_2)
\nonu \\
&+& (\mbox{Non-linear singular terms}) +\cdots,
\label{open2linear}
\eea
where the nonlinear terms are given in the original paper \cite{Ahn94}.
As we take the large $c$ limit blindly, 
the quadratic term has $\frac{1}{c}$-behavior and 
the cubic term has $\frac{1}{c^2}$-behavior. For the linear term in
(\ref{open2linear}),
all the fields in the right hand side 
behave like as $c$ independent term except the $[D, \overline{D}]
T(w)$ which goes to $\frac{1}{c}$. 
In particular, the self-coupling constant appearing in the right hand
side of (\ref{open2linear}) can be written as
\bea
\alpha_{N=2}^2=
\frac{(c+3)^2(5c-12)^2}{2 (15-c) (-1+c) (6+c) (-3+2 c)},
\nonu
\eea
which can be generalized to the $N=4$ (\ref{const2}) and the arbitrary
$N$ (\ref{selfcoupling}). 
It is also useful to compare the above expression (\ref{open2linear})
with 
the previous results in the component approach \cite{Romans}.

\section{ The generators and structure constants of $SU(5)$ in complex
  basis in the subsection \ref{n4} }

One can choose $4$ diagonal Cartan generators in the normalization of
\cite{Georgi}. 
Let us describe the $8$ generators $T_m$ where $m=1, 2, \cdots, 8$
and the remaining $4$ generators $T_a$ where $a=9, 10, 11, 12$ (\ref{indices}) as follows:
\bea
T_1 & = &
\left(
\begin{array}{ccccccc}
0 & 0 &0 &0 &0   \\ 
1 & 0 &0 &0 &0  \\
0 & 0 &0 &0 &0 \\
0 & 0 &0 & 0 & 0  \\
0 & 0 &0 & 0 & 0 \\
\end{array} \right),
T_2 =
\left(
\begin{array}{ccccccc}
0 & 0 &0 &0 &0   \\ 
0 & 0 &0 &0 &0  \\
1 & 0 &0 &0 &0 \\
0 & 0 &0 & 0 & 0  \\
0 & 0 &0 & 0 & 0 \\
\end{array} \right), 
T_3 =
\left(
\begin{array}{ccccccc}
0 & 0 &0 &0 &0   \\ 
0 & 0 &0 &0 &0  \\
0 & 0 &0 &0 &0 \\
1 & 0 &0 & 0 & 0  \\
0 & 0 &0 & 0 & 0 \\
\end{array} \right),
\nonu \\
T_4  & = &
\left(
\begin{array}{ccccccc}
0 & 0 &0 &0 &0   \\ 
0 & 0 &0 &0 &0  \\
0 & 1 &0 &0 &0 \\
0 & 0 &0 & 0 & 0  \\
0 & 0 &0 & 0 & 0 \\
\end{array} \right),
T_5 =
\left(
\begin{array}{ccccccc}
0 & 0 &0 &0 &0   \\ 
0 & 0 &0 &0 &0  \\
0 & 0 &0 &0 &0 \\
0 & 1 &0 & 0 & 0  \\
0 & 0 &0 & 0 & 0 \\
\end{array} \right), 
T_6 =
\left(
\begin{array}{ccccccc}
0 & 0 &0 &0 &0   \\ 
0 & 0 &0 &0 &0  \\
0 & 0 &0 &0 &0 \\
0 & 0 &1 & 0 & 0  \\
0 & 0 &0 & 0 & 0 \\
\end{array} \right),
\nonu \\
T_7  & = &
\left(
\begin{array}{ccccccc}
\frac{i}{2}+\frac{1}{\sqrt{12}} & 0 &0 &0 &0   \\ 
0 & -\frac{i}{2} +\frac{1}{\sqrt{12}} &0 &0 &0  \\
0 & 0 &-\frac{2}{\sqrt{12}} &0 &0 \\
0 & 0 &0 & 0 & 0  \\
0 & 0 &0 & 0 & 0 \\
\end{array} \right),
\nonu \\
T_8  &= &
\left(
\begin{array}{ccccccc}
\frac{i}{\sqrt{24}}+\frac{1}{\sqrt{40}}  & 0 &0 &0 &0   \\ 
0 & \frac{i}{\sqrt{24}}+\frac{1}{\sqrt{40}} &0 &0 &0  \\
0 & 0 &\frac{i}{\sqrt{24}}+\frac{1}{\sqrt{40}} &0 &0 \\
0 & 0 &0 & -\frac{3i}{\sqrt{24}}+\frac{1}{\sqrt{40}} & 0  \\
0 & 0 &0 & 0 & -\frac{4}{\sqrt{40}} \\
\end{array} \right), 
\nonu \\
T_9 & = &
\left(
\begin{array}{ccccccc}
0 & 0 &0 &0 &0   \\ 
0 & 0 &0 &0 &0  \\
0 & 0 &0 &0 &0 \\
0 & 0 &0 & 0 & 0  \\
1 & 0 &0 & 0 & 0 \\
\end{array} \right),
T_{10} =
\left(
\begin{array}{ccccccc}
0 & 0 &0 &0 &0   \\ 
0 & 0 &0 &0 &0  \\
0 & 0 &0 &0 &0 \\
0 & 0 &0 & 0 & 0  \\
0 & 1 &0 & 0 & 0 \\
\end{array} \right), 
T_{11} =
\left(
\begin{array}{ccccccc}
0 & 0 &0 &0 &0   \\ 
0 & 0 &0 &0 &0  \\
0 & 0 &0 &0 &0 \\
0 & 0 &0 & 0 & 0  \\
0 & 0 &1 & 0 & 0 \\
\end{array} \right),
\nonu \\
T_{12} & = &
\left(
\begin{array}{ccccccc}
0 & 0 &0 &0 &0   \\ 
0 & 0 &0 &0 &0  \\
0 & 0 &0 &0 &0 \\
0 & 0 &0 & 0 & 0  \\
0 & 0 &0 & 1 & 0 \\
\end{array} \right).
\nonu
\eea
Note that $T_7= i H_1 + H_2$ and $T_8 = i H_3 + H_4$ where the four
Cartan generators are given 
$H_1 = \frac{1}{2} \mbox{diag}(1,-1,0,0,0)$, $H_2 =
\frac{1}{\sqrt{12}} \mbox{diag}(1,1,-2,0,0)$, $H_3 =
\frac{1}{\sqrt{24}} 
\mbox{diag}(1,1,1,-3,0)$ and $H_4 = 
\frac{1}{\sqrt{40}} \mbox{diag}(1,1,1,1,-4)$ \cite{Georgi}.
The conjugated generators $T_{\bar{m}}$ and $T_{\bar{a}}$ can be
obtained from the relations from the Hermiticity $T_{\bar{m}} =T_m^{\dagger}$ and
$T_{\bar{a}}=T_a^{\dagger}$ as we mentioned before. 
One can easily see that the $14$ generators $T_m, T_{\bar{m}}$ where $m=1,
2, \cdots, 7$ and the $(T_8 -T_{\bar{8}})$ consist of 
the $SU(4)$ subgroup generators. The remaining diagonal generator 
$(T_8 +T_{\bar{8}})$ corresponds to the $U(1)$ in the subgroup $H$ of
the coset model.

Moreover, the $492$ nonzero 
structure constants can be obtained from the commutator relations
\bea
\left[ T_m, T_n \right] & = & f_{m n}^{\;\;\;\; p} T_p +   f_{m n}^{\;\;\;\;\bar{p}}
T_{\bar{p}},
\qquad 
\left[ T_m, T_{\bar{n}} \right] = 
f_{m \bar{n}}^{\;\;\;\; p} T_p +   f_{m \bar{n}}^{\;\;\;\;\bar{p}}
T_{\bar{p}},
\nonu \\
\left[ 
T_{\bar{m}}, T_{\bar{n}} 
\right] 
& = & 
f_{\bar{m} \bar{n}}^{\;\;\;\;p} T_p +
f_{\bar{m} \bar{n}}^{\;\;\;\;\bar{p}}
T_{\bar{p}},
\qquad
\left[ 
T_{m}, T_{a} 
\right] 
 = 
f_{m a}^{\;\;\;\;b} T_b,
\nonu \\
\left[ 
T_{\bar{m}}, T_{a} \right] 
& = & 
f_{\bar{m} a}^{\;\;\;\;b} T_b,
\qquad
\left[ 
T_{\bar{m}}, T_{\bar{a}} 
\right] 
 = 
f_{\bar{m} \bar{a}}^{\;\;\;\;\bar{b}} T_{\bar{b}},
\qquad
\left[
T_{m}, T_{\bar{a}} 
\right] 
 = 
f_{m \bar{a}}^{\;\;\;\;\bar{b}} T_{\bar{b}}.
\nonu
\eea

It is straightforward to obtain the complex basis for general
$N$. This will be necessary to analyze the ${\cal N}=2$ ${\cal
  W}_{N+1}$ algebra. It is nontrivial to write down two $U(1)$
generators in the subgroup $H$ explicitly.

\section{The coefficient functions 
in the primary field (\ref{superspin2}) in ${\cal N}=2$
${\cal W}_5$  algebra}

For the spin $2$ current (\ref{superspin2}), the coefficient functions
can be obtained from the two conditions (\ref{KWvanishing}) and (\ref{TW}):
\bea
b_1 &=& \frac{i (-25+k (23+8 k)) }{6 \sqrt{2} (5+k)} A(k),
\qquad
b_2  =  \frac{i (5-11 k) }{6 \sqrt{2}} A(k),
\nonu \\
b_3 & = & \frac{i (9+k) (-5+2 k)}{12 \sqrt{2}} A(k),
\qquad
b_4  =  -\frac{i k (9+k) }{4 \sqrt{2}} A(k),
\nonu \\
b_5 & = & \frac{i k (9+k) }{4 \sqrt{2}} A(k),
\qquad
b_6  =  \frac{i (-50+k (73+19 k)) }{12 \sqrt{2}} A(k),
\nonu \\
b_7 & = & \frac{i (-50+k (73+19 k)) }{12 \sqrt{2}} A(k),
\qquad
b_8  =  \frac{i k (9+k) }{4 \sqrt{2}} A(k),
\nonu \\
b_9 & = & -\frac{i k (5+k) (9+k) }{2 \sqrt{2}} A(k),
\qquad
b_{10}  =   \frac{i (5+k) (-5+11 k) }{6 \sqrt{2}} A(k),
\nonu \\
b_{11} & = & \frac{5 i (-20+k (-1+k (4+k))) }{8 \sqrt{2}} A(k),
\qquad
b_{12}  =   -\frac{5 i (-20+k (-1+k (4+k))) }{8 \sqrt{2}} A(k),
\nonu \\
b_{13} & = & -\frac{i (5+k) (-5+11 k)}{12 \sqrt{2}} A(k),
\qquad
b_{14}  =   -\frac{i (50+k (305+k (50+3 k))) }{24 \sqrt{2}} A(k),
\nonu \\
b_{15} & = & -\frac{i k (5+k) (9+k) }{8 \sqrt{2}} A(k),
\qquad
b_{16}  =  -\frac{i k (5+k) (9+k) }{8 \sqrt{2}} A(k),
\nonu \\
b_{17} & = & \frac{i (5+k) (-5+11 k) }{12 \sqrt{2}} A(k),
\qquad
b_{18}  =   -\frac{i (5+k) (10+k (5+3 k)) }{24 \sqrt{2}} A(k),
\nonu \\
b_{19} & = &  -\frac{i (5+k) (-5+11 k) }{6 \sqrt{2}} A(k),
\qquad
b_{20}  =   \frac{i (5+k)^2 (1+3 k) }{12 \sqrt{2}} A(k),
\nonu \\
b_{21} & = & \frac{i (5+k) (-5+11 k) }{12 \sqrt{2}} A(k),
\qquad
b_{22} =   \frac{i \left(-20+k+k^2\right)}{3 \sqrt{2}} A(k),
\nonu \\
b_{23} & = &  \frac{i \left(-20+k+k^2\right) }{3 \sqrt{2}} A(k),
\qquad
b_{24}  =  \frac{i (5+k) (35+k) }{60 \sqrt{2}} A(k),
\nonu \\
b_{25} & = & \frac{i (5+k) (35+k) }{30 \sqrt{2}} A(k),
\qquad
b_{26}  =  \frac{i (5+k) (35+k) }{60 \sqrt{2}} A(k),
\nonu \\
b_{27} & = & \frac{i k (5+k) (9+k)}{4 \sqrt{2}} A(k),
\label{coeffn5}
\eea
where the overall coefficient function $A(k)$ is determined and is
given by (\ref{Ak}).
One can also analyze the large $k$ limit where the $N$ is fixed along
the line of \cite{GS}. Then some of the coefficients in
(\ref{coeffn5}) can survive.

\section{The operator product expansions in the component approach 
for (\ref{final}) in ${\cal N}=2$ ${\cal W}_{N+1}$ algebra }

In practice, it is better to compute the operator product expansions
in the component approach(there are some problems in using the package
\cite{KT} directly) and they are given by
\bea
&& W(z) \; W(w)  =  
\nonu \\
&& \frac{1}{(z-w)^4} \frac{c(N,k)}{2} +\frac{1}{(z-w)^2} \left[
  2 \alpha(N,k) W -[D, \overline{D}] T \right](w)   \nonu \\
&& +  \frac{1}{(z-w)} \left[ 
\alpha(N,k) \pa W -\frac{1}{2} \pa [D, \overline{D}] T \right](w) 
 +   \cdots,
\label{ope345} \\
&& W(z) \; (D W+\overline{D} W)(w)  =  
\nonu \\
&&
\frac{1}{(z-w)^3} 3 \left[ D T -\overline{D} T \right](w)
 +\frac{1}{(z-w)^2} \left[
\alpha(N,k) (D W +\overline{D} W) +  \pa (D T -\overline{D} T) \right](w) 
\nonu \\
&& + \frac{1}{(z-w)} \left[ ( D V - \overline{D} V) +\frac{2}{5} \alpha(N,k)
\pa (D W +\overline{D} W) +\frac{1}{4} \pa^2 (D T -\overline{D} T)
\right](w) \nonu \\
&&
+  \frac{1}{N} \mbox{(Non-linear singular terms)} +\cdots,
\label{ope348} \\
&& W(z) \; (D W-\overline{D} W)(w)  =  
\nonu \\
&&
\frac{1}{(z-w)^3} 3 \left[ D T +\overline{D} T \right](w)
 +\frac{1}{(z-w)^2} \left[
\alpha(N,k) (D W - \overline{D} W) +  \pa (D T + \overline{D} T) \right](w)
\nonu \\
&& + \frac{1}{(z-w)} \left[ ( D V + \overline{D} V) +\frac{2}{5} \alpha(N,k)
\pa (D W -\overline{D} W) +\frac{1}{4} \pa^2 (D T +\overline{D} T)
\right](w) \nonu \\
&& +  \frac{1}{N} \mbox{(Non-linear singular terms)}+\cdots,
\label{ope349} \\
&& W(z) \; (-1)\frac{1}{2} [D, \overline{D}] W(w)  =  
\nonu \\
&&
\frac{1}{(z-w)^4} 3 T(w) +\frac{1}{(z-w)^2} \left[-\frac{3}{10} \alpha(N,k) [D, 
\overline{D} ] W + 3 V \right](w) 
\nonu \\
&& +  
\frac{1}{(z-w)} \left[ -\frac{1}{10} \alpha(N,k) \pa [D, 
\overline{D} ] W + \pa V \right](w) +  \frac{1}{N} \mbox{(Non-linear
singular terms)}
\nonu \\
&& +\cdots,
\label{ope346} 
\\
&& (D W+\overline{D} W)(z) \; W(w)  =  
\nonu \\
&&
-\frac{1}{(z-w)^3} 3 \left[ D T -\overline{D} T \right](w)
 +\frac{1}{(z-w)^2} \left[
\alpha(N,k) (D W +\overline{D} W) -  2 \pa (D T -\overline{D} T) \right](w) 
\nonu \\
&& + \frac{1}{(z-w)} \left[ -( D V - \overline{D} V) +\frac{3}{5} \alpha(N,k)
\pa (D W +\overline{D} W) -\frac{3}{4} \pa^2 (D T -\overline{D} T)
\right](w) \nonu \\
&& +  \frac{1}{N} \mbox{(Non-linear singular terms)}+\cdots,
\nonu \\
&& (D W-\overline{D} W)(z) \; W(w)  =  
\nonu \\
&&
-\frac{1}{(z-w)^3} 3 \left[ D T +\overline{D} T \right](w)
 +\frac{1}{(z-w)^2} \left[
\alpha(N,k) (D W - \overline{D} W) -2  \pa (D T + \overline{D} T) \right](w)
\nonu \\
&& + \frac{1}{(z-w)} \left[ -( D V + \overline{D} V) +\frac{3}{5} \alpha(N,k)
\pa (D W -\overline{D} W) -\frac{3}{4} \pa^2 (D T +\overline{D} T)
\right](w) \nonu \\
&& +  \frac{1}{N} \mbox{(Non-linear singular terms)}+\cdots,
\nonu \\
&& D W (z) \; D W (w) =+ \cdots,  \label{zero1} \\
&& \overline{D} W(z) \; \overline{D} W(w) = +\cdots,  \label{zero2} \\ 
&& D W(z) \; \overline{D} W (w) +\overline{D} W(z) \; D W(w)  =  
\nonu \\
&&
\frac{1}{(z-w)^5} 2 c(N,k) +\frac{1}{(z-w)^3} \left[ 4 \alpha(N,k) W -5 [D,
  \overline{D}]T 
\right](w) \nonu \\
&& +\frac{1}{(z-w)^2} \left[ 2 \alpha(N,k) \pa W - \frac{5}{2} \pa [D,
  \overline{D}] T 
\right](w) \nonu \\
&& +\frac{1}{(z-w)} \left[\frac{3}{5} \alpha(N,k) \pa^2 W -
  \frac{3}{4} \pa^2 [D, \overline{D}] T - [D, \overline{D}] V
\right](w) \nonu \\
&& +  \frac{1}{N} \mbox{(Non-linear singular terms)}+\cdots,
\label{ope351} \\
&& D W(z)  \; \overline{D} W (w) -\overline{D} W(z) \; D W(w)  =  
\nonu \\
&&
\frac{1}{(z-w)^4} 6 T(w) +\frac{1}{(z-w)^3} 3 \pa T(w)  
+\frac{1}{(z-w)^2} \left[ \frac{2}{5} \alpha(N,k) [D, \overline{D}] W + \pa^2 T +6 V 
\right](w) \nonu \\
&& +\frac{1}{(z-w)} \left[ \frac{1}{5} \alpha(N,k) \pa [D, \overline{D}] W +
  \frac{1}{4} \pa^3 T + 3 \pa V \right](w) \nonu \\
&& +  \frac{1}{N} \mbox{(Non-linear singular terms)} +\cdots,
\label{ope352}
\\
&& (D W +\overline{D} W)(z) \; (-1)\frac{1}{2} [D, \overline{D}] W(w)  =
\nonu \\
&&
\frac{1}{(z-w)^4} \frac{15}{2} \left[ DT +\overline{D} T \right](w)
+\frac{1}
{(z-w)^3} \left[ \alpha(N,k) (D W -\overline{D} W)  + \frac{5}{2} \pa ( D T
  +\overline{D} T) \right](w) \nonu \\
&& + \frac{1}{(z-w)^2} \left[ \frac{2}{5}\alpha(N,k) \pa (D W -\overline{D} W)
  + 
\frac{5}{8} \pa^2 ( D T
  +\overline{D} T) +\frac{7}{2} (D V + \overline{D} V) \right](w)
\nonu \\
&& +
\frac{1}{(z-w)} \left[  \frac{1}{10}\alpha(N,k) \pa^2 (D W -\overline{D} W)
  + 
\frac{1}{8} \pa^3 ( D T
  +\overline{D} T) + \frac{3}{2} \pa (D V + \overline{D} V)\right](w)
\nonu \\
&& +  \frac{1}{N} \mbox{(Non-linear singular terms)} +\cdots,
\nonu \\
&&  (D W -\overline{D} W)(z) \; (-1)\frac{1}{2} [D, \overline{D}] W(w) =
\nonu \\
&&
\frac{1}{(z-w)^4} \frac{15}{2} \left[ DT -\overline{D} T \right](w)
 +\frac{1}
{(z-w)^3} \left[ \alpha(N,k) (D W +\overline{D} W)  + \frac{5}{2} \pa ( D T
  -\overline{D} T) \right](w) \nonu \\
&& + \frac{1}{(z-w)^2} \left[ \frac{2}{5}\alpha(N,k) \pa (D W +\overline{D} W)
  + 
\frac{5}{8} \pa^2 ( D T
  -\overline{D} T) +\frac{7}{2} (D V - \overline{D} V) \right](w)
\nonu \\
&& +
\frac{1}{(z-w)} \left[  \frac{1}{10}\alpha(N,k) \pa^2 (D W + \overline{D} W)
  + 
\frac{1}{8} \pa^3 ( D T
  -\overline{D} T) + \frac{3}{2} \pa (D V - \overline{D} V)\right](w)
\nonu \\
&& +  \frac{1}{N} \mbox{(Non-linear singular terms)} +\cdots,
\nonu \\
&& (-1)\frac{1}{2} [D, \overline{D}] W(z) \; W(w)  =  
\nonu \\
&&
\frac{1}{(z-w)^4} 3 T(w) +\frac{1}{(z-w)^3} 3 \pa T(w)  + 
 \frac{1}{(z-w)^2} \left[-\frac{3}{10} \alpha(N,k) [D, 
\overline{D} ] W + 3 V +\frac{3}{2} \pa^2 T \right](w) 
\nonu \\
&& +  
\frac{1}{(z-w)} \left[ -\frac{1}{5} \alpha(N,k) \pa [D, 
\overline{D} ] W + 2 \pa V +\frac{1}{2} \pa^3 T \right](w) \nonu \\
&& +  \frac{1}{N} \mbox{(Non-linear singular terms)} +\cdots,
\nonu \\
&& (-1)\frac{1}{2} [D, \overline{D}] W(z) \; (D W +\overline{D} W)(w) =
\nonu \\
&&
\frac{1}{(z-w)^4} \frac{15}{2} \left[ DT +\overline{D} T \right](w)
+\frac{1}
{(z-w)^3} \left[ -\alpha(N,k) (D W -\overline{D} W)  + 5 \pa ( D T
  +\overline{D} T) \right](w) \nonu \\
&& + \frac{1}{(z-w)^2} \left[ -\frac{3}{5}\alpha(N,k) \pa (D W -\overline{D} W)
  + 
\frac{15}{8} \pa^2 ( D T
  +\overline{D} T) +\frac{7}{2} (D V + \overline{D} V) \right](w)
\nonu \\
&& +
\frac{1}{(z-w)} \left[  -\frac{1}{5}\alpha(N,k) \pa^2 (D W -\overline{D} W)
  + 
\frac{1}{2} \pa^3 ( D T
  +\overline{D} T) +2 \pa (D V + \overline{D} V)\right](w) \nonu \\
&& +  \frac{1}{N} \mbox{(Non-linear singular terms)}
+\cdots,
\nonu \\
&& (-1)\frac{1}{2} [D, \overline{D}] W(z) \; (D W -\overline{D} W)(w) =
\nonu \\
&&
\frac{1}{(z-w)^4} \frac{15}{2} \left[ DT -\overline{D} T \right](w)
 +\frac{1}
{(z-w)^3} \left[ -\alpha(N,k) (D W +\overline{D} W)  + 5 \pa ( D T
  -\overline{D} T) \right](w) \nonu \\
&& + \frac{1}{(z-w)^2} \left[ -\frac{3}{5}\alpha(N,k) \pa (D W +\overline{D} W)
  + 
\frac{15}{8} \pa^2 ( D T
  -\overline{D} T) +\frac{7}{2} (D V - \overline{D} V) \right](w)
\nonu \\
&& +
\frac{1}{(z-w)} \left[  -\frac{1}{5}\alpha(N,k) \pa^2 (D W + \overline{D} W)
  + 
\frac{1}{2} \pa^3 ( D T
  -\overline{D} T) +2 \pa (D V - \overline{D} V)\right](w) \nonu \\
&& +  \frac{1}{N} \mbox{(Non-linear singular terms)} +\cdots,
\label{ope350} \\
&& (-1)\frac{1}{2} [D, \overline{D}] W(z) \;
 (-1)\frac{1}{2} [D, \overline{D}] W(w) 
=
\nonu \\
&& \frac{1}{(z-w)^6} \frac{5}{2} c(N,k) +\frac{1}{(z-w)^4} \left[ 3 \alpha(N,k) W
  -\frac{15}{2} [D, \overline{D}] T\right](w) 
\nonu \\
&& + \frac{1}{(z-w)^3} \left[ \frac{3}{2} \alpha(N,k) \pa W
  -\frac{15}{4} \pa [D, \overline{D}] T \right](w) \nonu \\
&& +
\frac{1}{(z-w)^2} \left[ \frac{9}{20} \alpha(N,k) \pa^2 W
  -\frac{9}{8} \pa^2 [D, \overline{D}] T -2 [D, \overline{D}] V  \right](w)
\nonu \\
&& + \frac{1}{(z-w)} \left[  \frac{1}{10} \alpha(N,k) \pa^3 W
  -\frac{1}{4} \pa^3 [D, \overline{D}] T - \pa [D, \overline{D}] V
\right](w) 
\nonu \\
&& +  \frac{1}{N} \mbox{(Non-linear singular terms)} +\cdots,
\label{ope347}
\eea
where we write down the operator product expansions for the $D W(z)$ and
$\overline{D} W(z)$ by taking the sum or difference between them 
because we want to compare them with the asymptotic symmetry algebra
in the $AdS_3$ bulk theory directly. One can also obtain the separate
operator product expansions by adding or subtracting the corresponding
equations without any difficulty.

In section $4$, we only presented some of the large $(N,k)$ limit for
the full algebra. Here we complete them by repeating those analysis as
follows.
For the operator product expansion between the spin $2$ and spin
$\frac{5}{2}$, one has, from (\ref{ope348}), 
\bea
&& W_{2,HP}^{-}(z) G_{3,HP}^{-}(w)  =  
\frac{1}{(z-w)^3} 3 \beta(N,k)^2 G_{2,HP}^{+}(w)
\nonu \\
&& +\frac{1}{(z-w)^2}  \beta(N,k)^2 \left[
\frac{\alpha(N,k)}{\beta(N,k)} G_{3,HP}^{-} +  \pa  G_{2,HP}^{+} \right](w) 
\nonu \\
&& + \frac{1}{(z-w)}  \beta(N,k)^2 \left[ \frac{2}{5} \frac{\alpha(N,k)}{\beta(N,k)}
\pa G_{3,HP}^{-} +\frac{1}{4} \pa^2  G_{2,HP}^{+} + 
 \frac{1}{\beta(N,k)^2} G_{4,HP}^{+} \right](w) \nonu \\
&& +  \frac{1}{N} \mbox{(Non-linear singular terms)} +\cdots
\nonu \\
&& \longrightarrow 
\frac{1}{(z-w)^3}  (-1) \frac{2}{3} (2\la_{HP}+1)(\la_{HP}-1) G_{2,HP}^{+}(w) \nonu \\
&& + \frac{1}{(z-2)^2} \left[
 \frac{1}{3}(1-4\la_{HP})  G_{3,HP}^{-}- \frac{2}{9}
 (2\la_{HP}+1)(\la_{HP}-1) 
\pa G_{2,HP}^{+}  \right](w)
\nonu \\
&& + \frac{1}{(z-w)}  \left[
 \frac{2}{15}(1-4\la_{HP})  \pa G_{3,HP}^{-}- \frac{1}{18}
 (2\la_{HP}+1)(\la_{HP}-1) \pa^2
G_{2,HP}^{+} + G_{4,HP}^{+} \right](w)
\nonu \\
&& +  \frac{1}{N} \mbox{(Non-linear singular terms)} + \cdots.
\label{remain1}
\eea
This corresponds to the equation of $(3.48)$ of \cite{HP}
\footnote{The commutator is given by $ [(W_{2,HP}^{-})_m,
  (G_{3,HP}^{-})_n]= \beta(N,k)^2 \frac{1}{16} \left(-9+12 m^2-8 m n+4
    n^2\right)
 (G_{2,HP}^{+})_{m+n} + \beta(N,k)^2
 \frac{\alpha(N,k)}{5\beta(N,k)}(3m-2n)
(G_{3,HP}^{-})_{m+n} +(G_{4,HP}^{+})_{m+n}+\mbox{Nonlinear terms}$.}.
We also use the normalizations like as (\ref{relations}).
Similarly, one has the following operator product expansion 
between the spin $2$ current and other spin $\frac{5}{2}$ current,
from (\ref{ope349}),
\bea
&& W_{2,HP}^{-}(z) G_{3,HP}^{+}(w)  =  
\frac{1}{(z-w)^3} 3 \beta(N,k)^2 G_{2,HP}^{-}(w)
\nonu \\
&& +\frac{1}{(z-w)^2}  \beta(N,k)^2 \left[
\frac{\alpha(N,k)}{\beta(N,k)} G_{3,HP}^{+} +  \pa  G_{2,HP}^{-} \right](w) 
\nonu \\
&& + \frac{1}{(z-w)}  \beta(N,k)^2 \left[ \frac{2}{5} \frac{\alpha(N,k)}{\beta(N,k)}
\pa G_{3,HP}^{+} +\frac{1}{4} \pa^2  G_{2,HP}^{-} + 
 \frac{1}{\beta(N,k)^2} G_{4,HP}^{-} \right](w) \nonu \\
&& +  \frac{1}{N} \mbox{(Non-linear singular terms)} +\cdots
\nonu \\
&& \longrightarrow 
\frac{1}{(z-w)^3}  (-1) \frac{2}{3} (2\la_{HP}+1)(\la_{HP}-1) G_{2,HP}^{-}(w) \nonu \\
&& + \frac{1}{(z-2)^2} \left[
 \frac{1}{3}(1-4\la_{HP})  G_{3,HP}^{+}- \frac{2}{9}
 (2\la_{HP}+1)(\la_{HP}-1) 
\pa G_{2,HP}^{-}  \right](w)
\nonu \\
&& + \frac{1}{(z-w)}  \left[
 \frac{2}{15}(1-4\la_{HP})  \pa G_{3,HP}^{+}- \frac{1}{18}
 (2\la_{HP}+1)(\la_{HP}-1) \pa^2
G_{2,HP}^{-} + G_{4,HP}^{-} \right](w)
\nonu \\
&& +  \frac{1}{N} \mbox{(Non-linear singular terms)} + \cdots,
\label{remain2}
\eea
which agrees with the equation $(3.49)$ of \cite{HP} \footnote{This
is equivalent to $ [(W_{2,HP}^{-})_m,
  (G_{3,HP}^{+})_n]= \beta(N,k)^2 \frac{1}{16} \left(-9+12 m^2-8 m n+4
    n^2\right)
 (G_{2,HP}^{-})_{m+n} + \beta(N,k)^2
 \frac{\alpha(N,k)}{5\beta(N,k)}(3m-2n)
(G_{3,HP}^{+})_{m+n} +(G_{4,HP}^{-})_{m+n}+\mbox{Nonlinear terms}$.}.

One has the following operator product expansion between the spin $3$
current and the spin $\frac{5}{2}$ current:
\bea
&& W_{3,HP}^{+}(z) G_{3,HP}^{-}(w) =
\frac{1}{(z-w)^4} \frac{15}{2} \beta(N,k)^2 G_{2,HP}^{-}(w)
\nonu \\
&& +\frac{1}
{(z-w)^3}  \beta(N,k)^2 \left[ -\frac{\alpha(N,k)}{\beta(N,k)} 
G_{3,HP}^{+}  + 5 \pa G_{2,HP}^{-} \right](w) \nonu \\
&& + \frac{1}{(z-w)^2}  \beta(N,k)^2 \left[ -\frac{3}{5}
  \frac{\alpha(N,k)}{\beta(N,k)} 
\pa  G_{3,HP}^{+} 
  + 
\frac{15}{8} \pa^2 G_{2,HP}^{-}  +\frac{7}{2}  \frac{1}{\beta(N,k)^2} G_{4,HP}^{-}  \right](w)
\nonu \\
&& +
\frac{1}{(z-w)}  \beta(N,k)^2 \left[  -\frac{1}{5} \frac{\alpha(N,k) }
{\beta(N,k)} \pa^2  G_{3,HP}^{+}
  + 
\frac{1}{2} \pa^3 G_{2,HP}^{-} +2 \frac{1}{\beta(N,k)^2} \pa
G_{4,HP}^{-} \right](w) 
\nonu \\
&& +  \frac{1}{N} \mbox{(Non-linear singular terms)} +\cdots
\nonu \\
&& \longrightarrow \frac{1}{(z-w)^4}  (-1) \frac{5}{3}
(2\la_{HP}+1)(\la_{HP}-1)  G_{2,HP}^{-}(w) 
\nonu \\
&& + \frac{1}{(z-w)^3}  \left[ 
 -\frac{1}{3}(1-4\la_{HP})  G_{3,HP}^{+} 
  - \frac{10}{9}
(2\la_{HP}+1)(\la_{HP}-1)
  \pa G_{2,HP}^{-}  \right]
\nonu \\
&& + \frac{1}{(z-w)^2}  \left[ 
 -\frac{1}{5}(1-4\la_{HP})  \pa G_{3,HP}^{+} 
  - \frac{5}{12}
(2\la_{HP}+1)(\la_{HP}-1)
  \pa^2 G_{2,HP}^{-}  +\frac{7}{2}  G_{4,HP}^{-}\right]
\nonu \\
&& + \frac{1}{(z-w)}  \left[ 
 -\frac{1}{15}(1-4\la_{HP}) \pa^2 G_{3,HP}^{+} 
  - \frac{1}{9}
(2\la_{HP}+1)(\la_{HP}-1)
  \pa^3 G_{2,HP}^{-}  + 2  \pa G_{4,HP}^{-}\right]
\nonu \\
&& +  \frac{1}{N} \mbox{(Non-linear singular terms)} + \cdots.
\label{remain3}
\eea
One can also write down the commutator \footnote{That is,
$[(W_{3,HP}^{+})_m, (G_{3,HP}^{-})_n]=\frac{1}{32} 
\left(-19 m+4 m^3+36 n-8 m^2 n+12 m n^2-16 n^3\right) \beta(N,k)^2  
(G_{3,HP}^{-})_{m+n} -\frac{1}{20} \left(-5+2 m^2-4 m n+4 n^2\right)
\alpha(N,k)  
\beta(N,k)(G_{3,HP}^{+})_{m+n} +\left( \frac{3}{2}m -
  2n\right)(G_{4,HP}^{-})_{m+n}+
\mbox{Nonlinear terms}$.}.

Furthermore, one obtains, from (\ref{ope350}) and the normalization (\ref{defV}), 
\bea
&& W_{3,HP}^{+}(z) G_{3,HP}^{+}(w) =
\frac{1}{(z-w)^4} \frac{15}{2} \beta(N,k)^2 G_{2,HP}^{+}(w)
\nonu \\
&& +\frac{1}
{(z-w)^3}  \beta(N,k)^2 \left[ -\frac{\alpha(N,k)}{\beta(N,k)} 
G_{3,HP}^{-}  + 5 \pa G_{2,HP}^{+} \right](w) \nonu \\
&& + \frac{1}{(z-w)^2}  \beta(N,k)^2 \left[ -\frac{3}{5}
  \frac{\alpha(N,k)}{\beta(N,k)} 
\pa  G_{3,HP}^{-} 
  + 
\frac{15}{8} \pa^2 G_{2,HP}^{+}  +\frac{7}{2}  \frac{1}{\beta(N,k)^2} G_{4,HP}^{+}  \right](w)
\nonu \\
&& +
\frac{1}{(z-w)}  \beta(N,k)^2 \left[  -\frac{1}{5} \frac{\alpha(N,k) }
{\beta(N,k)} \pa^2  G_{3,HP}^{-}
  + 
\frac{1}{2} \pa^3 G_{2,HP}^{+} +2 \frac{1}{\beta(N,k)^2} \pa
G_{4,HP}^{+} \right](w) 
\nonu \\
&& +  \frac{1}{N} \mbox{(Non-linear singular terms)} +\cdots
\nonu \\
&& \longrightarrow \frac{1}{(z-w)^4}  (-1) \frac{5}{3}
(2\la_{HP}+1)(\la_{HP}-1)  G_{2,HP}^{+}(w) 
\nonu \\
&& + \frac{1}{(z-w)^3}  \left[ 
 -\frac{1}{3}(1-4\la_{HP})  G_{3,HP}^{-} 
  - \frac{10}{9}
(2\la_{HP}+1)(\la_{HP}-1)
  \pa G_{2,HP}^{+}  \right]
\nonu \\
&& + \frac{1}{(z-w)^2}  \left[ 
 -\frac{1}{5}(1-4\la_{HP}) \pa G_{3,HP}^{-} 
  - \frac{5}{12}
(2\la_{HP}+1)(\la_{HP}-1)
  \pa^2 G_{2,HP}^{+}  +\frac{7}{2}  G_{4,HP}^{+}\right]
\nonu \\
&& + \frac{1}{(z-w)}  \left[ 
 -\frac{1}{15}(1-4\la_{HP}) \pa^2 G_{3,HP}^{-} 
  - \frac{1}{9}
(2\la_{HP}+1)(\la_{HP}-1)
  \pa^3 G_{2,HP}^{+}  + 2  \pa G_{4,HP}^{+}\right]
\nonu \\
&& +  \frac{1}{N} \mbox{(Non-linear singular terms)} + \cdots,
\label{remain4}
\eea
which will lead to the equation $(3.50)$ of \cite{HP}
\footnote{That is, $[(W_{3,HP}^{+})_m, (G_{3,HP}^{+})_n]=\frac{1}{32} 
\left(-19 m+4 m^3+36 n-8 m^2 n+12 m n^2-16 n^3\right) \beta(N,k)^2  
(G_{3,HP}^{+})_{m+n} -\frac{1}{20} \left(-5+2 m^2-4 m n+4 n^2\right)
\alpha(N,k)  
\beta(N,k)(G_{3,HP}^{-})_{m+n} +\left( \frac{3}{2}m -
  2n\right)(G_{4,HP}^{+})_{m+n}+
\mbox{Nonlinear terms}$. }.

The operator product expansion of spin $\frac{5}{2}$ and itself, from
(\ref{ope351}) with (\ref{zero1}) and (\ref{zero2}), 
leads to
\bea
&& G_{3,HP}^{-}(z) G_{3,HP}^{-}(w)  =  
\frac{1}{(z-w)^5} 2 c(N,k) \beta(N,k)^2 
\nonu \\
&& +\frac{1}{(z-w)^3}  \beta(N,k)^2  \left[ 4 \frac{\alpha(N,k)
  }{\beta(N,k)} 
W_{2,HP}^{-} +10 W_{2,HP}^{+} 
\right](w) \nonu \\
&& +\frac{1}{(z-w)^2}  \beta(N,k)^2 \left[ 2
  \frac{\alpha(N,k)}{\beta(N,k)} 
\pa W_{2,HP}^{-} +5 \pa W_{2,HP}^{+} 
\right](w) \nonu \\
&& +\frac{1}{(z-w)}  \beta(N,k)^2 \left[\frac{3}{5} 
\frac{\alpha(N,k)}{\beta(N,k)} \pa^2 W_{2,HP}^{-} +
  \frac{3}{2} \pa^2 W_{2,HP}^{+}  +\frac{2}{\beta(N,k)^2} W_{4,HP}^{+}
\right](w) 
\nonu \\
&& +  \frac{1}{N} \mbox{(Non-linear singular terms)} +\cdots
\nonu \\
&& \longrightarrow 
\frac{1}{(z-w)^5} (-1)\frac{4}{3}(1-\la_{HP})(2\la_{HP}-1)(2\la_{HP}+1)N
\nonu \\
&& + \frac{1}{(z-w)^3} \left[ \frac{4}{3}(1-4\la_{HP}) W_{2,HP}^{-} +
\frac{20}{9}(1-\la_{HP})(2\la_{HP}+1) W_{2,HP}^{+}  \right](w)
\nonu \\
&& + \frac{1}{(z-w)^2} \left[ \frac{2}{3}(1-4\la_{HP}) \pa W_{2,HP}^{-} +
\frac{10}{9}(1-\la_{HP})(2\la_{HP}+1) \pa W_{2,HP}^{+}  \right](w)
\nonu \\
&& + \frac{1}{(z-w)} \left[ \frac{1}{5}(1-4\la_{HP}) \pa^2 W_{2,HP}^{-} +
\frac{1}{3}(1-\la_{HP})(2\la_{HP}+1) \pa^2 W_{2,HP}^{+}
+2 W_{4,HP}^{+}
  \right](w)
\nonu \\
&& +  \frac{1}{N} \mbox{(Non-linear singular terms)} + \cdots,
\label{remain5}
\eea
which agrees with the equation $(3.51)$ of \cite{HP} \footnote{Also in
  this case one
has the following anticommutator 
$ \{(G_{3,HP}^{-})_m, (G_{3,HP}^{-})_n \}= \frac{1}{12} c(N,k)
\beta(N,k)^2 (m^2-\frac{1}{4})(m^2-\frac{9}{4})+ \frac{1}{10} 
\left(-9+6 m^2-8 m n+6 n^2\right) \left[ 
\alpha(N,k)  \beta(N,k) (W_{2,HP}^{-})_{m+n} +\frac{5}{2}
(W_{2,HP}^{+})_{m+n} \right] + 2(W_{4,HP}^{+})_{m+n} +\mbox{Nonlinear
terms}$.}. 
We also use the
similar normalization for $[D, \overline{D}] V(w) $ as (\ref{defV}).
Furthermore, one has, from (\ref{ope352}) with (\ref{zero1}) and (\ref{zero2}), 
\bea
&& G_{3,HP}^{-}(z) G_{3,HP}^{+}(w)  =  
-\frac{1}{(z-w)^4} 6 \beta(N,k)^2 W_{1,HP}^{-}(w) 
-\frac{1}{(z-w)^3} 3 \beta(N,k)^2 \pa W_{1,HP}^{-}(w)  \nonu \\
&& -\frac{1}{(z-w)^2} \beta(N,k)^2 \left[ -\frac{4}{5} \frac{\alpha(N,k)}{\beta(N,k)} 
W_{3,HP}^{+} + \pa^2 W_{1,HP}^{-} + \frac{6}{\beta(N,k)^2} W_{3,HP}^{-} 
\right](w) \nonu \\
&& -\frac{1}{(z-w)} \beta(N,k)^2 \left[ -\frac{2}{5} \frac{\alpha(N,k)
  }{\beta(N,k)} \pa 
W_{3,HP}^{+} +
  \frac{1}{4} \pa^3 W_{1,HP}^{-} + \frac{3}{\beta(N,k)^2} \pa
  W_{3,HP}^{-} \right](w) 
\nonu \\
&& +  \frac{1}{N} \mbox{(Non-linear singular terms)} +\cdots
\nonu \\
&& \longrightarrow \frac{1}{(z-w)^4}  
  \frac{4}{3} (2\la_{HP}+1)(\la_{HP}-1)  W_{1,HP}^{-}(w) 
\nonu \\
&& +\frac{1}{(z-w)^3}  \frac{2}{3} (2\la_{HP}+1)(\la_{HP}-1)  \pa W_{1,HP}^{-}(w) 
\nonu \\
&& + \frac{1}{(z-w)^2} \left[  \frac{4}{15}(1-4\la_{HP}) W_{3,HP}^{+}
  +
 \frac{2}{9} (2\la_{HP}+1)(\la_{HP}-1)  \pa^2 W_{1,HP}^{-}  -
 6 W_{3,HP}^{-} 
 \right](w)
\nonu \\
&& + \frac{1}{(z-w)} \left[  \frac{2}{15}(1-4\la_{HP}) \pa W_{3,HP}^{+}
  +
 \frac{1}{18} (2\la_{HP}+1)(\la_{HP}-1)  \pa^3  W_{1,HP}^{-}  -
 3 \pa W_{3,HP}^{-} 
 \right](w)
\nonu \\
&& +  \frac{1}{N} \mbox{(Non-linear singular terms)} + \cdots,
\label{remain6}
\eea
which coincides with the equation $(3.52)$ of \cite{HP} and we also
use the normalization like (\ref{defV}) \footnote{There is also 
an anticommutator relation for the fermionic modes as follows:
$\{  (G_{3,HP}^{-})_m,  (G_{3,HP}^{+})_n \}=-\frac{1}{8} 
\left(24+39 m+24 m^2+6 m^3+49 n+48 m n+10 m^2 n+24 n^2+14 m n^2+2
  n^3\right) 
\beta(N,k)^2 (W_{1,HP}^{-})_{m+n} + (m-n) \left[ \frac{2}{5}
  \alpha(N,k) \beta(N,k)  (W_{3,HP}^{+})_{m+n} - 3(W_{3,HP}^{-})_{m+n}  
\right]+ \mbox{Nonlinear terms}$. }.
Similar analysis can be done for the operator product expansions
(\ref{ope345})-(\ref{ope347}) 
we did
not consider.


\end{document}